\documentclass[twocolumn]{aastex631}

\usepackage{amsmath}
\usepackage{dsnr-2024sep}
\usepackage{savesym}
\savesymbol{tablenum}
\usepackage{siunitx}
\restoresymbol{SIX}{tablenum}
\usepackage{siunitx}
\usepackage{xcolor}

\newcommand{\hst}{\emph{HST}}

\DeclareSIUnit\photon{photon}
\DeclareSIUnit\count{counts}
\DeclareSIUnit\pixel{pixel}
\DeclareSIUnit\kpc{kpc}
\DeclareSIUnit\pc{pc}
\DeclareSIUnit\erg{erg}
\DeclareSIUnit\arcsec{arcsec}

\received{21 Dec 2024}
\revised{22 Mar 2025}
\accepted{TBD}

\submitjournal{AAS Journals}

\shorttitle{\osix\ Imaging of Makani}
\shortauthors{Ha et al.}

\graphicspath{{./}{figures/}}

\begin{document}

\title{Deep Ultraviolet, Emission-Line Imaging of the Makani Galactic Wind}

\correspondingauthor{David Rupke}
\email{drupke@gmail.com}

\author[0009-0000-3788-9059]{Triet Ha}
\affiliation{Department of Physics, Rhodes College, 2000 N. Parkway, Memphis, TN 38104, USA}

\author[0000-0002-1608-7564]{David S. N. Rupke}
\affiliation{Department of Physics, Rhodes College, 2000 N. Parkway, Memphis, TN 38104, USA}
\affiliation{Zentrum für Astronomie der Universität Heidelberg, Astronomisches Rechen-Institut, Mönchhofstr 12-14, D-69120 Heidelberg, Germany}

\author{Shane Caraker}
\affiliation{Department of Physics, Rhodes College, 2000 N. Parkway, Memphis, TN 38104, USA}

\author{Jack Harper}
\affiliation{Department of Physics, Rhodes College, 2000 N. Parkway, Memphis, TN 38104, USA}

\author[0000-0002-2583-5894]{Alison L. Coil}
\affil{Department of Astronomy and Astrophysics, University of California, San Diego, La Jolla, CA 92093, USA}

\author[0000-0003-0773-582X]{Miao Li}
\affil{Flatiron Institute, Center for Computational Astrophysics, New York, NY 10010, USA}

\author[0000-0003-1809-6920]{Christy A. Tremonti}
\affil{Department of Astronomy, University of Wisconsin-Madison, Madison, WI 53706, USA}

\author{Aleksandar M. Diamond-Stanic}
\affil{Department of Physics and Astronomy, Bates College, Lewiston, ME, 04240, USA}

\author{James E. Geach}
\affil{Centre for Astrophysics Research, University of Hertfordshire, Hatfield, Hertfordshire AL10 9AB, UK}

\author[0000-0003-1468-9526]{Ryan C. Hickox}
\affil{Department of Physics and Astronomy, Dartmouth College, Hanover, NH 03755, USA}

\author[0000-0001-9487-8583]{Sean D. Johnson}
\affil{Department of Astronomy, University of Michigan, 1085 S. University, Ann Arbor, MI 48109, USA}

\author[0000-0002-9393-6507]{Gene C. K. Leung}
\affil{MIT Kavli Institute for Astrophysics and Space Research, 77 Massachusetts Ave., Cambridge, MA 02139, USA}

\author[0000-0002-2733-4559]{John Moustakas}
\affil{Department of Physics and Astronomy, Siena College, Loudonville, NY 12211, USA}

\author[0000-0002-2451-9160]{Serena Perrotta}
\affil{Department of Astronomy and Astrophysics, University of California, San Diego, La Jolla, CA 92093, USA}

\author[0000-0001-5851-1856]{Gregory H. Rudnick}
\affil{Department of Physics and Astronomy, University of Kansas, Lawrence, KS 66045, USA}

\author[0000-0003-1771-5531]{Paul H. Sell}
\affil{Department of Astronomy, University of Florida, Gainesville, FL, 32611 USA}

\author[0000-0002-8571-9801]{Kelly E. Whalen}
\affil{NASA Goddard Space Flight Center, Code 662, Greenbelt, 20771, MD, USA}

\begin{abstract}
The \osixl\ line is a key probe of cooling gas in the circumgalactic medium (CGM) of galaxies, but has been observed to date primarily in absorption along single sightlines. We present deep \hst\ ACS-SBC observations of the compact, massive starburst Makani. Makani hosts a 100~kpc, \ot-emitting galactic wind driven by two episodes of star formation over 400~Myr. We detect \osix\ and \lya\ emission across the \ot\ nebula with similar morphology and extent, out to $r\approx \qty{50}{\kpc}$. Using differential narrow-band imaging, we separate \lya\ and \osix\ and show that the \osix\ emission is comparable in brightness to \ot, with $L_\mathrm{OVI} =\qty{4e42}{\erg\per\second}$. The similar hourglass morphology and size of \ot\ and \osix\ implicate radiative cooling at $T = 10^{5.5}$~K in a hot--cold interface. This may occur as the $T > 10^7$~K CGM---or the hot fluid driving the wind---exchanges mass with the $T \approx 10^4$~K clouds entrained in (or formed by) the wind. The optical/UV line ratios may be consistent with shock ionization, though uncertain attenuation and \lya\ radiative transfer complicate the interpretation. The detection of \osix\ in Makani lies at the bleeding edge of the UV imaging capabilities of \hst, and provides a benchmark for future emission-line imaging of the CGM with a wide-area UV telescope.
\end{abstract}

%% Keywords should appear after the \end{abstract} command. 
%% The AAS Journals now uses Unified Astronomy Thesaurus concepts:
%% https://astrothesaurus.org
%% You will be asked to selected these concepts during the submission process
%% but this old "keyword" functionality is maintained in case authors want
%% to include these concepts in their preprints.
%\keywords{Hot air}

\section{Introduction} \label{sec:intro}

The circumgalactic medium (CGM) surrounds galaxies and fills their virial halos with hot, diffuse, and metal-enriched gas.  These metals are ejected by galactic winds from central and satellite galaxies \citep{2011Sci...334..952T,2019MNRAS.488.1248H}. The CGM likely contains the majority of metals produced in galaxies \citep{2011Sci...334..948T, 2014ApJ...792....8W}.

Much of the mass in the CGM may reside in a warm-hot phase that traces cooling gas \citep{2014ApJ...792....8W}. A key tracer of this phase is \ion{O}{6}~1031.912,~1037.613~\AA\ in absorption, whose emissivity peaks in a narrow range around $T\approx10^{5.5}$~K \citep{1993ApJS...88..253S}. Quasars or galaxies back-illuminate the CGM of galaxies that are nearby in projection. The statistical analysis of \osix\ absorbers shows that they are most common around actively star-forming galaxies \citep{2011Sci...334..948T, 2023ApJ...949...41T}, suggesting a connection between the ionization state of oxygen in the CGM and the feedback process.

Imaging the CGM in emission is much more challenging, as many of the expected line coolants lie in the UV and X-ray bands \citep{2013MNRAS.430.3292B}. \osixl\ is one of the most observable of these---due to its brightness and that it arises conveniently in wavelength near \lya---despite it not being the dominant ionization state of oxygen in the CGM \citep{2016MNRAS.460.2157O}. Presently, deep, narrowband observations are still required to image extended \osix. Current UV-sensitive telescopes (primarily {\em HST}) are hard-pressed to meet the sensitivity required to detect the diffuse, low-surface-brightness emission that is expected.

Using two {\it Far Ultraviolet Spectroscopic Explorer} ({\it FUSE}) spectra, \citet{2003ApJ...591..821O} were the first to detect \osix\ in emission from a spiral galaxy other than the Milky Way. The \osix\ emission in NGC~4631 has a scale height of \qty{8}{\kpc} and is coincident with an outflowing, X-ray emitting bubble along the galaxy minor axis. Weak \osix\ emission was later detected with {\it FUSE} in integrated spectra of nearby star-forming galaxies \citet{2007ApJ...668..891G, 2009ApJS..181..272G}.

A synthetic narrowband imaging technique was successfully deployed by \citet{2016ApJ...828...49H} to image the \ion{O}{6} emission surrounding the nearby galaxy SDSS J115630.63$+$500822.1 (hereafter J1156) using the Solar Blind Channel of the Advanced Camera for Surveys (ACS-SBC) aboard {\em HST}. They detect \osix\ emission out to \qty{23}{\kpc} in an exponential halo with scale length $r_\mathrm{exp} = \qty{7.5}{\kpc}$ and luminosity $L_\mathrm{OVI} = \qty{2e41}{\erg\per\second}$. By comparing to column densities measured in absorption, \citet{2016ApJ...828...49H} infer that the emission arises from the interface between hot gas and small (\qty{<10}{\pc}), cool clouds \citep[see also][]{2018MNRAS.474.1688C} in a starburst-driven galactic wind.

The giant Makani nebula is one of the first examples of a galactic wind that is directly observed to be moving well into the CGM of its host \citep{2019Natur.574..643R}. The nebula was detected in the light of \otl\ and has an observed luminosity $L(\ot)=\qty{3.3e42}{\erg\per\second}$. Its radius is 50~kpc, which is 25$\times$ the stellar half-light radius of the host galaxy. The hourglass morphology and gas kinematics indicate the nebula arises from an outflow. The host galaxy---SDSS J211824.06+001729.4, a.k.a. Makani---is a $z=0.4590$, $M_* = 10^{11}~\msun$ merger remnant hosting a compact starburst \citep{2014MNRAS.441.3417S,2021ApJ...912...11D}. With a star formation rate of $225-300$~\smpy\ \citep{2020ApJ...901..138P} and a fast wind with $(dM/dt)/\mathrm{SFR}\sim1$ \citep{2021ApJ...923..275P,2023ApJ...947...33R}, it is a likely example of an Eddington-limited starburst \citep{2012ApJ...755L..26D}.

The wind consists of a two-stage flow. First, a $\langle \sigma \rangle = \qty{200}{\km\per\second}$ ionized outflow at radii \qtyrange{20}{50}{\kpc} emerged from a starburst $\approx$400~My in the past (Episode~I). Second, a  $\langle \sigma \rangle = \qty{400}{\km\per\second}$, neutral/molecular/ionized flow---with velocities up to \qty{2000}{\km\per\second}---was driven out to \qtyrange{10}{20}{\kpc} by a recent burst of age 7~Myr (Episode~II). Makani is part of a sample of mergers \citep{2014MNRAS.441.3417S} that may be progressing through a brief stage \citep{2022AJ....164..222W} of stellar mass buildup \citep{2021ApJ...912...11D} driven by starbursts that power high-velocity \citep{2023ApJ...951..105D}, massive \citep{2023ApJ...949....9P} outflows.

The size and brightness of the Makani oxygen nebula make it a prime candidate for imaging \ion{O}{6} in a galactic wind as it interacts with the CGM. The rest-frame optical lines and molecular gas emission are consistent with a massive (10$^{10}$~\msun), shock-ionized wind \citep{2019Natur.574..643R,2023ApJ...947...33R}. \citet{2016ApJ...828...49H} find that the coronal, \osix-emitting gas in J1156 arises at much larger scales than the compact, photoionized line emission. Determining where the \osix\ arises in Makani, and its relationship to the line emission from lower-ionization gas, will inform our understanding of the relationshsip between cold clouds in a wind and the hot gas in which they are embedded. This interaction is a subject of intense interest, as it impacts our understanding of the origin and fate of the clouds driven in a galactic wind, and the redistribution of gas and metals into the CGM \citep{2022ApJ...924...82F}.

In Section~\ref{sec:obs-red} we lay out the observations and data reduction. In Section~\ref{sec:results} we present the resulting observed morphology, interpret the source of the observed emission, and compute line fluxes. Finally, we compare to simulations and models in Section~\ref{sec:discuss}. Throughout we take as systemic the stellar redshift of Makani, $z=0.4590$ \citep{2019Natur.574..643R} and assume a flat $\Lambda$ cosmology with $\Omega_m=0.315$ and $H_0=67.4$~\kms~Mpc$^{-1}$ \citep{2020A&A...641A...6P}. This results in a projected physical scale of \qty{6.02}{\kpc\per\arcsec} at the distance of Makani. The vacuum rest wavelengths of the lines of interest in this paper are \qty{1031.912}{\text{\AA}} (\osix~\qty{1032}{\text{\AA}}), \qty{1037.613}{\text{\AA}} (\osix~\qty{1038}{\text{\AA}}) and \qty{1215.6701}{\text{\AA}} (Ly$\alpha$).

\clearpage

\begin{figure}[ht!]
\includegraphics[width=\columnwidth]{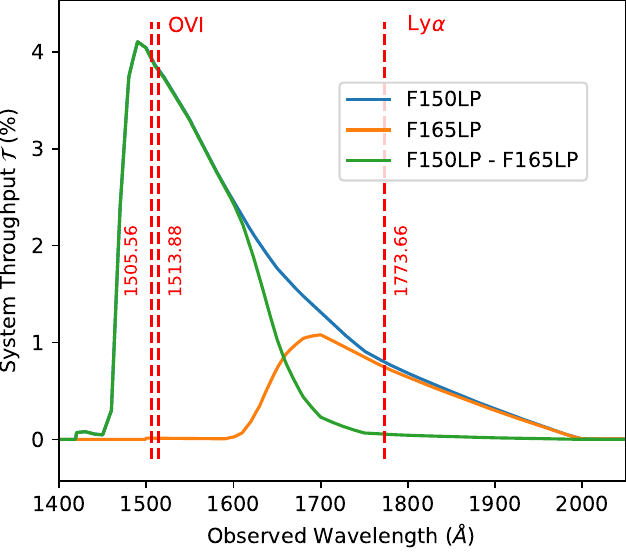}
\caption{System throughputs for the ACS-SBC F150LP and F165LP filters through which we imaged Makani \citep{2018ascl.soft11001S, 2020ascl.soft10003S, 2023acsi.book...23R}, shown as blue and orange lines; their difference is in green. The wavelengths of \osix\ and Ly$\alpha$ are marked as vertical, red, dashed lines, and labeled with their expected wavelengths in the observed frame. The difference of the two filters includes \osix\ emission but minimal Ly$\alpha$.}
\label{fig:throughputs}
\end{figure}

\section{Observations and Data Reduction} \label{sec:obs-red}
\subsection{Observations with HST} \label{sec:obs}

We imaged Makani with the ACS-SBC \citep{2003SPIE.4854..686T, 2005PASP..117.1049S} on {\em HST} as part of a Cycle 28 program (PID 16231, PI Rupke). We observed Makani through the F150LP and F165LP long-pass UV filters.

F150LP passes wavelengths above \qty{\approx 1450}{\text{\AA}} \citep{2018ascl.soft11001S, 2020ascl.soft10003S, 2023acsi.book...23R}, with a blue leak of throughput $\mathcal{T} < \qty{0.1}{\percent}$ down to \qty{1420}{\text{\AA}} and a tail to longer wavelengths that dips below $\mathcal{T} = \qty{0.01}{\percent}$ at \qty{2000}{\text{\AA}}. Figure \ref{fig:throughputs} illustrates the proximity of redshifted \osix\ wavelengths (\qty{1505}{\text{\AA}} and \qty{1514}{\text{\AA}}) to the wavelength of the peak of the system throughput for F150LP, \qty{1490}{\text{\AA}}. The throughput at redshifted \osix\ is \qtyrange{3.8}{3.9}{\percent}, compared to $\mathcal{T} = \qty{4.1}{\percent}$ at the filter peak. This is why Makani and F150LP are well-matched for this experiment, despite the overall low system throughput at these wavelengths.

F150LP also includes redshifted \lyb\ and \lya\ at \qty{1497}{\text{\AA}} and \qty{1774}{\text{\AA}}. We expect \lyb\ to be weak, as we discuss in Section~\ref{sec:discuss}. The throughput at \lya\ is $\mathcal{T} = \qty{0.79}{\percent}$, or about 5$\times$ smaller than the throughput near redshifted \osix. \lya\ is nonetheless a strong line, and could contaminate any measurement with F150LP alone.

F165LP passes wavelengths above \qty{\approx 1600}{\text{\AA}}, with a peak throughput of $\mathcal{T} = \qty{1.1}{\percent}$ at \qty{1700}{\text{\AA}} and a tail to longer wavelengths. F165LP looks much like F150LP, but with a higher low-wavelength cutoff. Notably, $\mathcal{T}_\mathrm{F150LP} - \mathcal{T}_\mathrm{F165LP} < \qty{0.1}{\percent}$ at wavelengths above \qty{1750}{\text{\AA}}. F165LP thus includes \lya, as well, but crucially not \osix. The difference of these two filters thus cleanly includes \osix\ but not \lya. The percent difference between the filters at redshfited \lya\ is only $\Delta\mathcal{T}/\mathcal{T} = \qty{7}{\percent}$. Thus, if \osix\ and \lya\ were equal in flux in Makani and no other lines were present, an image created by subtracting F165LP directly from F150LP would in principle contain residual \lya\ at \qty{2}{\percent} of the flux of \osix.
Neither filter contains strong geocoronal lines or significant amounts of Earthshine or Zodiacal light \citep{2023acsi.book...23R}.

We planned 20 orbits to reach a total exposure time of \qty{\approx 25000}{s} per filter. We grouped these into 10 visits of 2 orbits each. In each orbit we observed the galaxy in either the F150LP or F165LP filter in a multiple of the default 4-point box dither to subsample at the \onehalf\;pixel level. (We expanded the default dither by a factor of 5 to prevent overlap of the bad anode---the central blank strip of 6 rows \citep{2023acsi.book...23R}---among exposures.) In the second orbit of each visit, we switched filters and repeated the 4-point dither. At each dither step we exposed for \qty{624}{s} in orbit 1 and \qtyrange{626}{627}{s} in orbit 2. We alternated the first filter to be observed from visit to visit. This results in four exposures per filter per visit, for a total exposure time per visit of \qtyrange{2496}{2508}{s} in each filter.

The most significant noise source for the ACS-SBC MAMA detector, aside from photon statistics, is the dark current. For detector temperatures below \qty{\approx 25}{^\circ C}, the dark current is relatively stable at \qty{1e-5}{\count\per\second\per\pixel} \citep{2009acs..rept....2C,2017acs..rept....4A}. However, once the detector is powered on, its temperature increases with time---reaching \qty{>25}{^\circ C} after \qty{\approx 2}{hours} \citep{2017acs..rept....4A}---which in turn causes the dark current to rise \citep{2009acs..rept....2C}. Furthermore, there is a region of high dark current in a large, off-center ``halo'' on the detector. Thus, recommended practice is to limit ACS-SBC visits to 1--2 orbits; to keep ACS-SBC visits separated by at least 24 hours; and to place targets near the corner of the detector with lowest dark current \citep{2018acs..rept....7A}. The dark current remains relatively stable and low in this region at $T<\qty{25}{^\circ C}$, even at recent times \citep{2024acs..rept....NX}.

We followed the first two of these recommendations. Our visits were limited to 2 orbits, and we requested the ACS-SBC be switched off at least 24 hours between visits. Centering in the low-dark region was impossible, however. The long axis of the Makani nebula is \qty{17}{\arcsecond} in size, whereas the detector field of view (FOV) is \qtyproduct{34.6 x 30.5}{\arcsecond}. Furthermore, we dithered in a 9-point box pattern between visits---using a \qty{\approx 3.2}{\arcsecond} step size in each dimension---to average over low-frequency flatfield variations. We also did not specify the \hst\ roll angle, to maximize scheduling flexibility. Because we did not specify a particular orientation, the actual visit-to-visit grid rotated over time.

These constraints meant that the galaxy was typically positioned somewhere between the nominal \texttt{SBC-LODARK} aperture \citep{2018acs..rept....7A} and the detector center. However, we did shift the observing center toward the \texttt{SBC-LODARK} aperture as much as possible, while ensuring the full nebula would remain in the FOV.

In practice, a series of issues required multiple re-observations. The originally planned visits are labeled 01--10. (1) Visits 02 and 05 failed due to \hst\ safing in 2021 June. These were replaced by Visits 52 and 55. (2) These visits were scheduled but  did not occur due to a second safing in 2021 October. Visits 52 and 55 were then replaced by Visits 56 and 57. (3) Three orbits acquired no data due to failure of guide star reaquisition: orbit 2 of Visits 06, 07, and 59 (see below). Orbit 2 of Visits 06 and 07 were replaced with Visit 58. Orbit 2 of Visit 59 was not repeated. (4) Visit 09 was scheduled directly after Visit 08, so the dark current continued to increase and the noise in these data are high. (Visit 10 also occurred 19 hours after Visit 09---less than the optimal 24 hours---but the data are not severely impacted.) Visit 59 was implemented to replace Visit 09.

\subsection{Data Reduction} \label{sec:red}

\begin{figure}[ht!]
\includegraphics[width=\columnwidth]{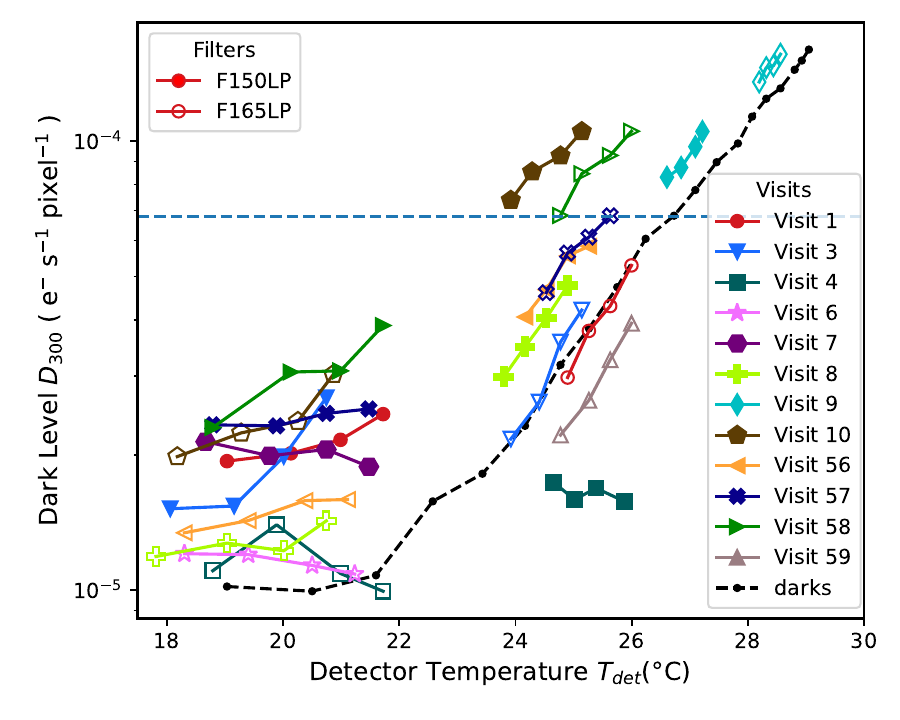}
\caption{The relationship between the average dark level $D_{300}$ and the detector temperature $T_\mathrm{det}$. $D_{300}$ is calculated in a \qtyproduct{300x300}{\pixel} box near the detector center, with the galaxy masked. The solid (empty) colored dots represent single flat-fielded exposures collected with the filter F150LP (F165LP). Different symbols denote different visits. The black dots represent data taken in 2022 with the camera shutter closed. Our data follow the 2022 dark pattern of increasing dark with increasing temperature above \qty{\approx20}{\degreeCelsius}, but with a varying normalization. The horizontal, blue, dashed line represents the cutoff used to eliminate data points with significantly high dark levels.}
\label{fig:dark}
\end{figure}

We retrieved 96 flat-fielded exposures---the \texttt{flt.fits} files---from MAST. Removing 12 with no data (orbit 2 of three visits in which guide star reacquisition failed) leaves 84 total exposures, for a total exposure time of \qty{52500}{\second}.
% (\dataset[10.17909/XXXXX]{http://dx.doi.org/10.17909/XXXXXX}).

We investigated the relationship between the dark current $D_{300}$ and the detector temperature $T_\mathrm{det}$ for each flat-fielded exposure (Figure~\ref{fig:dark}). We approximated $T_\mathrm{det}$ as the average values of the temperatures measured before and after the exposure, recorded in the header keywords \texttt{MDECODT1} and \texttt{MDECODT2}. The data was converted to units of electrons during the flat-field correction process \citep{2022acsd.book...11L}. For each file, we used its DQ array to exclude low-quality pixels and divided the data by the exposure time $t_{exp}$ obtained from the header keyword \texttt{EXPTIME}. These steps followed the methods discussed in \citet{2024acs..rept....NX}. We computed the dark level $D_{300}$ for each image by averaging pixel values within a \qtyproduct{300 x 300}{\pixel} square centered on the pixel (567, 561) while excluding a \qtyproduct{100 x 100}{\pixel} square around the galaxy center. This method ensured that the dark current measurement included the information from the highest dark current while avoiding the galaxy itself and the lowest dark current near the field edges. We retained $D_{300}$ to weight each image during registration. 

\begin{figure*}[ht!]
\includegraphics[width=\textwidth]{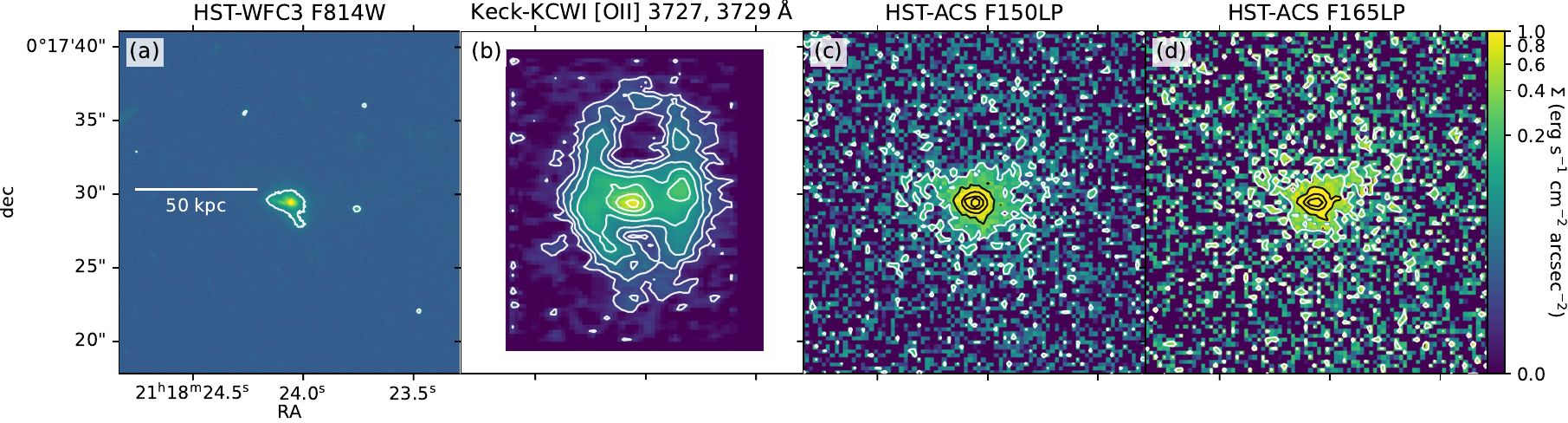}
\caption{\hst\ optical, Keck \ot, and \hst\ UV images. (a) \hst-WFC3 F814W image \citep{2014MNRAS.441.3417S}. The contour is 0.5~e$^-$\,\unit{\per\second\per\pixel}. The compact size of the galaxy is evident; half of the starlight is concentrated within a \qty{2}{\kpc} radius. (b) Makani in the light of \otl, as observed with Keck-KCWI \citep{2019Natur.574..643R}. Contours run from \qty{3.125e-18}{\erg\per\second\per\cm\squared\per\arcsec\squared} to \qty{2e-16}{\erg\per\second\per\cm\squared\per\arcsec\squared} and are spaced by factors of 2. The 100~kpc \ot\ nebula is $\approx$10$\times$ the size of the galaxy. (b) F150LP image of Makani, drizzled to match the KCWI pixel scale (\qtyproduct{0.2914 x 0.2914}{\arcsecond}). Contours are spaced 4$\times$ apart from \qty{3.125e-17}{\erg\per\second\per\cm\squared\per\arcsec\squared} to \qty{8e-15}{\erg\per\second\per\cm\squared\per\arcsec\squared}. We estimate the surface brightness by multiplying the count rate by \texttt{PHOTFLAM} and \texttt{PHOTBW} and dividing by pixel area in arcsec$^{-2}$. (c) Matching F165LP image, with contours down to \qty{6.25e-17}{\erg\per\second\per\cm\squared\per\arcsec\squared}. Some extended emission is evident in the uniformly-binned ACS-SBC images.}
\label{fig:img1}
\end{figure*}

At low temperatures ($T_\mathrm{det} < \qty{24}{^\circ C}$), the dark noise was uniformly flat at less than \qty{4e-5}{\count\per\second\per\pixel}. As the temperature rose, the dark level rose and the off-center halo became more prominent. This trend was discussed extensively in the previous ACS-SBC dark rate reports \citep{2009acs..rept....2C, 2017acs..rept....4A, 2024acs..rept....NX}. For comparison, we downloaded 20 SBC pre-calibrated dark images collected in March 2022 (PID 16529, PI Robert Avila). These 1000-second exposures were used by the ACS Team to monitor the dark rate of ACS-SBC and how it changed over time \citep{2024acs..rept....NX}. We performed CALACS calibration on these exposures to the point of flat-fielded correction. For each processed dark image, we computed the dark level per pixel, excluding bad-quality pixels flagged by the DQ array, and normalizing with the exposure time. We plot their dark levels versus the corresponding detector temperatures in Figure~\ref{fig:dark}.

Using these dark measurements, we eliminated two orbits in each filter with significantly high dark levels: Visit 09 orbit 1 and Visit 10 orbit 2 for F150LP; and Visit 09 orbit 2 and Visit 58 orbit 2 for F165LP. This cut left 36 exposures in F150LP and 32 in F165LP.

Prior to image registration, we corrected the shifts in coordinate zero-points caused by the use of different guiding stars in different visits. To do this, we adjusted the World Coordinate System (WCS) information in each exposure, matching the galaxy's centroid coordinate in each exposure with a reference coordinate. 

We drizzled images from the same filter onto a common grid using \texttt{DrizzlePac} \citep{2012ascl.soft12011S}. We passed our data to \texttt{astrodrizzle}, choosing an output plate scale of \qty{0.2914}{\arcsecond\per\pixel} to match the KCWI \ot\ image for a direct comparison. Since the output pixels are 9$\times$ bigger than the SBC detector pixels in each dimension, the pixel drop size---determined by the \texttt{final\_pixfrac} parameter---does not significantly affect the final image. Therefore, we chose $\texttt{final\_pixfrac}=1$ to minimize dark noise. During drizzling, we weighted each image by the ratio of its exposure time and dark level, $\mathbf{t_{exp}/D_{300}}$ \citep{2006AJ....132..853T}. We carried out the weighting process by first dividing each image and its exposure time by the dark level computed previously. Then, we combined these adjusted images with \texttt{astrodrizzle}, setting the parameter $\texttt{final\_wht\_type}=\texttt{EXP}$.

We obtained the variance map by passing the unweighted flat-fielded and WCS-shifted data again to \texttt{astrodrizzle}, but this time with $\texttt{final\_wht\_type}=\texttt{ERR}$. When individual exposures are processed, CALACS computes the noise $N$ from the signal $S$ as $N=\mathrm{max}(1,\sqrt{S})$, which accounts for the fact that most pixels record 0 counts \citep{2022acsd.book...11L}. Since we drizzle to much larger pixels, this resulted in an overestimate of the drizzled variance by a factor equal to the ratio of the original and drizzled pixel areas, which in this case was \qty{\approx 83}{}. We correct the variance map for this effect.

Finally, we computed the photometric centroid of each drizzled image and assigned to it a reference coordinate. To compute the reference coordinate, we downloaded the \hst-WFC3 F814W image \citep{2014MNRAS.441.3417S}, as reprocessed using the Hubble Advanced Pipeline Single-Visit Mosaics and aligned to Gaia DR3. We applied \texttt{reproject} \citep{2020ascl.soft11023R} to the WFC3 image to resample to an integer multiple of the KCWI platescale: $\qty{0.036425}{\arcsecond\per\pixel} = \qty{0.2914}{\arcsecond\per\pixel}/8$. This facilitates precisely co-registering the KCWI \ot\ image. We then computed the coordinate of the centroid of the resulting WFC3 reprojection. The result is $\mathrm{RA}=\ang[angle-symbol-over-decimal]{319.600245}=21\mathrm{h}18\mathrm{m}24.06\mathrm{s}$, $\delta=\ang[angle-symbol-over-decimal]{0.29151}=\ang[angle-symbol-over-decimal]{00;17;29.44}$.

\begin{figure*}[ht!]
\includegraphics[width=\textwidth]{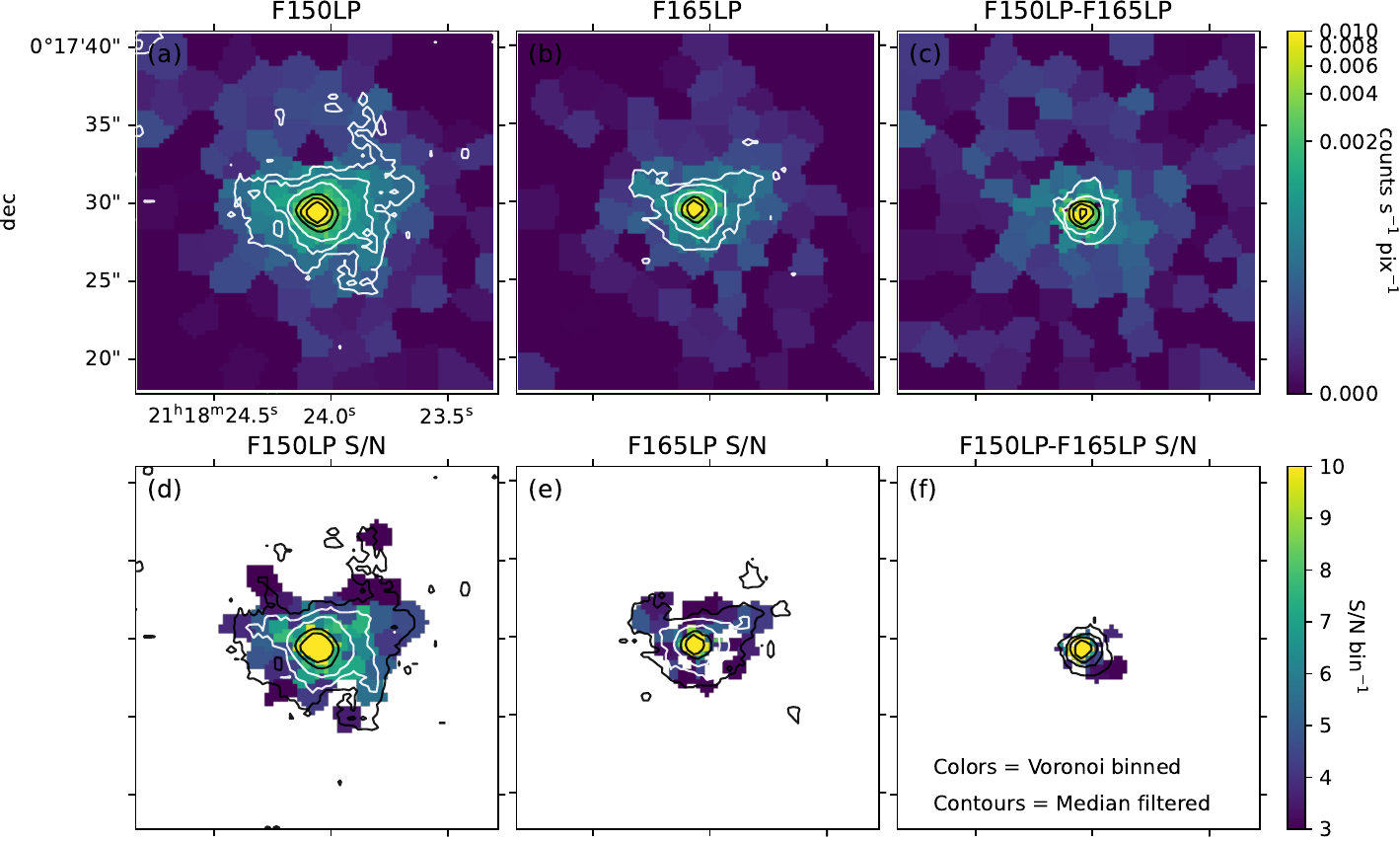}
\caption{(a)--(c) Images of F150LP, F165LP, and the difference $\mathrm{F150LP}-\mathrm{F165LP}$, respectively. Colors show Voronoi-binned images, while contours display uniformly-binned images which are median-filtered with a 5-pixel box. The colorbar at top right maps colors to counts~s$^{-1}$~pix$^{-1}$, while contours range from \qty{1e-2}{\count\per\second\per\pixel} down to \qty{3.125e-4}{} or \qty{6.25e-4}{} and are separated by factors of two. (d)--(e) Same as the top row, except S/N rather than intensity. Colors show S/N per Voronoi bin, as mapped by the colorbar at bottom right. Bins with $\mathrm{S/N}<3$ are not shown. Contours display S/N per pixel in the median-filtered images and associated error. They are separated by factors of two, ranging from 10 down to 0.6125 or 1.25.}
\label{fig:img2}
\end{figure*}

We thus obtained a WCS-aligned image and variance map in each filter with pixel scale \qtyproduct{0.2914 x 0.2914}{\arcsecond}. We show these images alongside the WFC3 and \ot\ images in Figure~\ref{fig:img1}, after dark subtaction (see below). We make an initial surface brightness estimate for each pixel by multiplying the count rate by the \texttt{PHOTFLAM} and \texttt{PHOTBW} values listed in the image headers and dividing by the area of each pixel in arcsec$^{-2}$. Some extended emission is already evident.

\begin{figure*}[ht!]
\includegraphics[width=\textwidth]{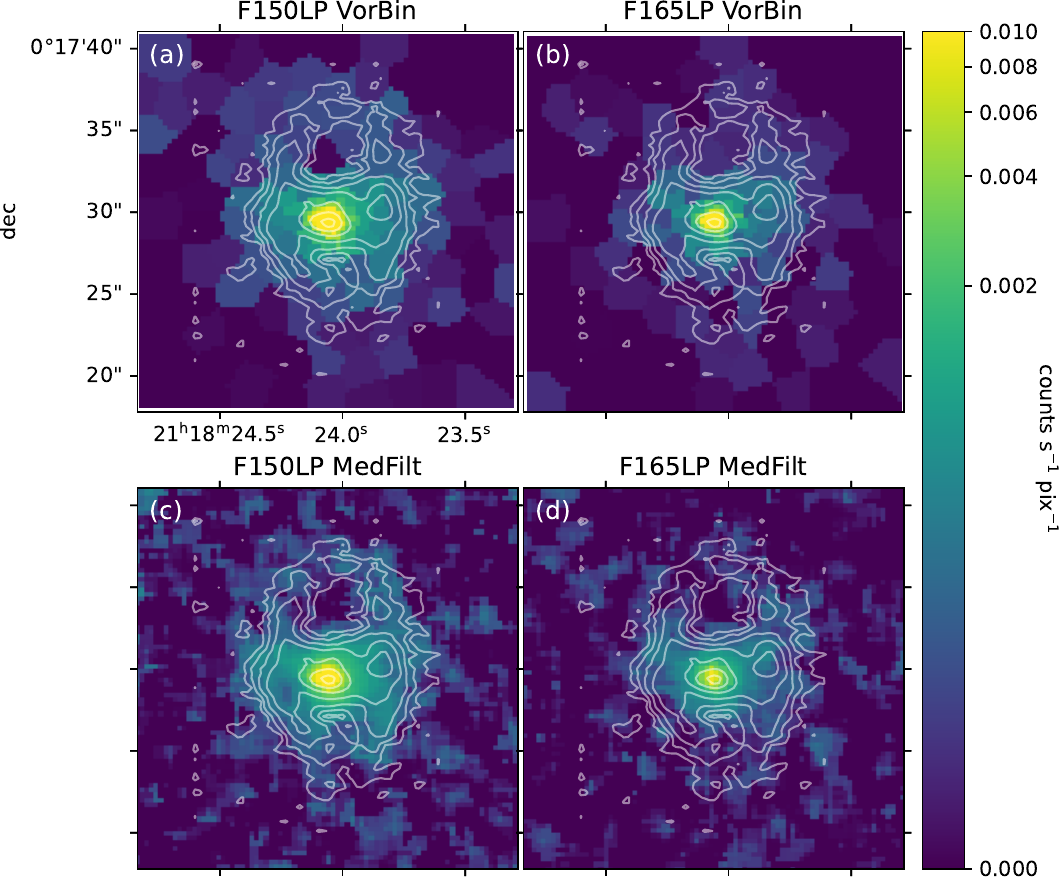}
\caption{A comparison of \ot\ and UV morphologies. (a) Voronoi-binned F150LP with \ot\ contours overlaid. The Voronoi image is identical to Figure~\ref{fig:img2}(a). Contours are the same as in Figure~\ref{fig:img1}(a). (b) Voronnoi-binned F165LP with \ot\ contours. (c) \ot\ on median-smoothed F150LP. (d) Smoothed F165LP, \ot\ contours. These images show the close morphological correspondence between F150LP and \ot. This correspondence is less prominent in F165LP, likely due to dark noise.}
\label{fig:o2-on-uv}
\end{figure*}

Since we eliminated exposures with high dark noise and inversely weighted each image by the dark level before drizzling, we expected the combined image to have a spatially uniform dark pattern that fills the FOV. In practice, in each image we found a mild curvature in the dark current with a shallow peak near or underneath the expected signal from the galaxy. This gradient results from the overlapping dark current halo structures, as they fill a much larger area on the detector than the galaxy. To subtract the dark noise, we then masked out the astrophysical structure in each image in an oval region. This structure was visible in each image prior to subtraction. As this structure is similar to that seen in the light of \ot, as discussed below, we used the \ot\ image as a reference during the masking process because of its higher signal-to-noise ratio ($S/N$). We also masked those pixels whose values were more than 2.5$\sigma$ away from the median in each image. We then fit the unmasked data points in each image with a two-dimensional polynomial. 2D polynomials of degree 3 and 5 were used as background models for the F150LP image and F165LP images, respectively. We subtracted the model from each image to create a background-subtracted image for each filter with pixel scale \qtyproduct{0.2914 x 0.2914}{\arcsecond}. The F150LP background averages \qty{1e-4}{\count\per\second\per\pixel} over the region of detected astrophysical signal---see Section \ref{sec:morph}---so that the Poisson noise from the galaxy is greater than the dark current Poisson noise over the detection region. However, the F165LP dark current is higher, ranging from \qty{5e-4}{\count\per\second\per\pixel} in the center down to \qty{2e-4}{\count\per\second\per\pixel} at the edge of the observed signal (see below in Section~\ref{sec:morph} for further discussion). Thus the dark current dominates the noise budget for the F165LP data at radii $r \ga \qty{2.5}{\arcsecond}$.

As the observed signal is very faint we further binned these images to enhance their $S/N$.  We applied Voronoi tesselation to the background-subtracted F150LP image with a target $\mathrm{S/N}\ga7$ \citep{2003MNRAS.342..345C}. During the binning process we weighted each pixel by $\omega_i =  |S_i/N_i^2|$, which maximizes $S/N$ while eliminating the contribution of pixels with significant noise and no signal \citep{1986PASP...98.1220R, 2003MNRAS.342..345C}. Since \osix\ lies at the peak of the F150LP throughput curve (Figure~\ref{fig:throughputs}), we used the resulting Voronoi bins from F150LP to bin other images to a common set of spatial bins. These include the F165LP images, a difference image, and the \ot\ image.

In Figure~\ref{fig:img2} we show the Voronoi-binned F150LP, F165LP, and $\mathrm{F150LP}-\mathrm{F165LP}$ images; and the $S/N$ per bin of each of these images Prior to Voronoi binning, we also median-filter each image with a 5-pixel box and overlay these contours in each panel of Figure~\ref{fig:img2} for comparison. As we discuss below, these Voronoi-binned and median-filtered images more fully reveal the extent of the UV emission.

\begin{figure*}[ht!]
\includegraphics[width=\textwidth]{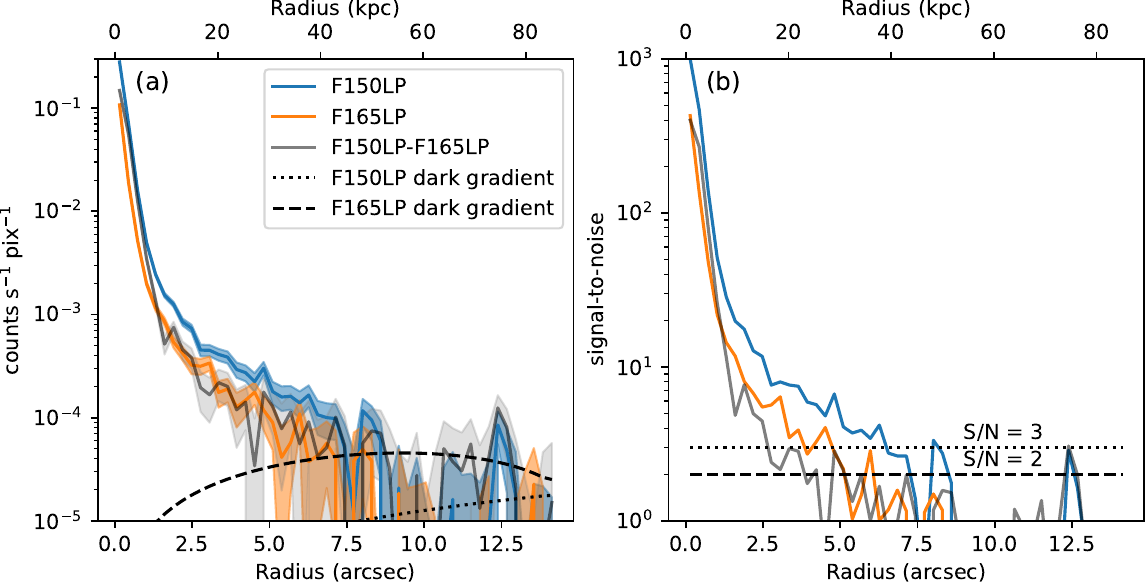}
\caption{(a) Radial profiles of F150LP and F165LP images from Figure~\ref{fig:img1}, as well as that of a difference image. 1$\sigma$ errors are shown as shaded regions. The F150LP and F165LP profiles are similar in shape, but there is more flux in the F150LP image at all radii. The difference profile is most similar to the F165LP profile. The dashed and dotted lines are the radial gradients in the dark models, and illustrate their potential contribution to the systemic uncertainty. The units of these gradients are \unit{\count\per\second\per\pixel\per\arcsec}. (b) S/N of radial profiles. Dashed and dotted lines denote $\mathrm{S/N}=2$ and 3 thresholds. These illustrate that the F165LP image only shows a significant detection at radii $r\la\qty{5}{\arcsecond}$, and that the difference image is even less informative.}
\label{fig:rad-prof-cts}
\end{figure*}

\section{Results} \label{sec:results}

\subsection{Morphology and Extent} \label{sec:morph}

The F150LP and F165LP images (Fig. 4) of Makani show a strongly-peaked central source. At a spatial scale of \qtyproduct{0.025 x 0.025}{\arcsecond}, the UV emission resolves into a luminous core, a clumpy elongation to the east, and a diffuse halo. These structures extend to a radius of \qty{\approx 1}{\arcsecond}, and are similar, though not identical, to those seen in the rest-frame optical. \citet{2014MNRAS.441.3417S} fit an $n=4$ S\'{e}rsic and point source model to the \hst\ F814W image (rest-frame $V$ band) of Makani (Figure~\ref{fig:img1}(a)). This leaves a 10\%\ flux residual, primarily in two tidal tails to the east and southwest. We leave the details of the $r < \qty{1}{\arcsecond}$ UV emission to future work but include in this section a few salient points. When binned to match the KCWI pixel scale of \qtyproduct{0.2914 x 0.2914}{\arcsecond}, the eastern elongation is visible in the centermost pixels in both filters (Figure~\ref{fig:img1}(c)--(d)). 

At radii of \qtyrange{1}{5}{\arcsecond}, UV emission extends away from the center to the east and west, in the same directions as the highest surface brightness \ot\ emission. The median-filtered and Voronoi-binned images reveal this emission most clearly (Figure~\ref{fig:img2}(a)--(b)). We show a direct comparison in Figure~\ref{fig:o2-on-uv}. We also plot azimuthally-averaged radial profiles of the drizzled images and their difference in Figure~\ref{fig:rad-prof-cts}(a) and a S/N radial profile of each in panel (b).

The F165LP image has its highest surface-brightness extensions \qty{\approx 5}{\arcsecond} to the northeast and northwest, with fainter emission to smaller radii in other directions (Figure~\ref{fig:img2}(b) and (e)). These NE and NW extensions are at the bases of the two lobes of the northern half of the \ot\ hourglass (Figure~\ref{fig:o2-on-uv}(b) and (d)). The radial S/N profile confirms that the F165LP emission is only detected above 2--3$\sigma$ at $r < \qty{5}{\arcsecond}$ (Figure~\ref{fig:rad-prof-cts}(b)).

The F150LP emission is much more extended and bears a strong resemblance to the \ot\ morphology. The Voronoi-binned data show 3$\sigma$ detections out to $r = \qty{8}{\arcsecond}$ (Figure~\ref{fig:img2}(d)). The radial profile confirms azimuthally-averaged emission at 2--3$\sigma$ out to $r \approx \qty{8.5}{\arcsecond}$ (Figure~\ref{fig:rad-prof-cts}(b)). The radius of the \ot\ nebula is also \qtyrange{8}{9}{\arcsecond} \citep{2019Natur.574..643R}. The F150LP emission is present in all four lobes of the \ot\ hourglass structure, as seen in both the Voronoi-binned and median-smoothed data (Figure~\ref{fig:o2-on-uv}(a) and (c)).

Considering also the data at lower significance (down to $S/N\approx2$), the F150LP data fully outline the cavity seen in the northern half of the \ot\ hourglass (Figure~\ref{fig:o2-on-uv}(a) and (c)). Though the southern half of the hourglass does not show a completely enclosed void in \ot, an emission minimum between two lobes is evident in the UV, as it is in \ot.

The difference image, $\mathrm{F150LP}-\mathrm{F165LP}$, shows significant emission only at $r < \qty{2.5}{\arcsecond}$ (Figure~\ref{fig:rad-prof-cts}(b)). This emission is not symmetric but is slightly extended to the west (Figure~\ref{fig:img2}(c) and (f)).

As we discuss above (Section~\ref{sec:obs}), the emission in F150LP contains the light through the F165LP filter plus emission from \qtyrange{1450}{\approx1600}{\text{\AA}} (Figure~\ref{fig:throughputs}). Additionally, the throughput of F150LP in the non-overlap region is up to 4$\times$ higher than in F165LP. As expected, the count rate per pixel $C$ is thus higher in F150LP than F165LP at all radii (Figure~\ref{fig:rad-prof-cts}(a)).

The radial profiles of F150LP, F165LP, and $\mathrm{F150LP}-\mathrm{F165LP}$ all show a similar decline with increasing radius (Figure~\ref{fig:rad-prof-cts}(a)). The F165LP and $\mathrm{F150LP}-\mathrm{F165LP}$ images decline to $C = \qty{1e-4}{\count\per\second\per\pixel}$ at \qty{\approx5}{\arcsecond}. The F150LP profile declines to this count rate at \qtyrange{\approx7}{8}{\arcsecond}. The noise floor in these images at about this level arises largely due to Poisson noise from dark counts. In particular, the combined dark level subtracted from the F165LP image is $\approx3\times$ higher than that of the F150LP image over the \qtyrange{0}{10}{\arcsecond} range. Thus, even though there are more astrophysical counts in the F150LP image, which in turn increases the Poisson noise, the noise is higher in our F165LP image due to increased dark current. The Poisson noise in the difference image is, of course, higher than in both.

As we discuss above, the radial significance cutoffs of each image are \qty{8.5}{\arcsecond} for F150LP, \qty{5}{\arcsecond} for F165LP, and \qty{2.5}{\arcsecond} for $\mathrm{F150LP}-\mathrm{F165LP}$. We attribute this in part to the decline in sensitivity across these images, but we also discuss astrophysical reasons for the difference below (Section~\ref{sec:source}).

A further source of potential error is uncertainty in the shape of the dark. In Figure~\ref{fig:rad-prof-cts}(a), we show the gradient in the dark model for each image. For F165LP in particular, this gradient reaches \qty{-4e-5}{\count\per\second\per\pixel\per\arcsec} at $r > \qty{7}{\arcsecond}$. The dark model is quite smooth, so this has no impact on discrete features such as those seen in Figure~\ref{fig:img2}. However, all three radial profiles appear to continue to decline slightly beyond their significance cutoffs. This could be in part due to model errors in the dark profile at these radii. Alternatively, it may be due to the low-count Poisson statistics that dominate these regions with little or no astrophysical signal.

\subsection{Origin of Emission} \label{sec:source}

We expect the regions within the central \qty{\approx 1}{\arcsecond} to be dominated by UV continuum and/or \lya\ emission, since this is where starlight and star formation in the system are concentrated. The compact starlight in Makani is evident from optical photometry. Makani has a  half-light radius of 0\farcs4 in the \hst\ F814W WFC3 image \citep{2014MNRAS.441.3417S}, though this significantly overestimates the size of the compact core due to the tidal features to the east and southwest(Section~\ref{sec:morph}). These tidal features extend to $r = \qty{2}{\arcsecond}$---within which 98\%\ of the rest-frame $V$-band light is contained, according to circular aperture photometry of the F814W image (Figure~\ref{fig:img1}(a)). Tractor photometry of deep, ground-based DESI Legacy Imaging Survey DR10.1 data of Makani \citep{2016ascl.soft04008L, 2019AJ....157..168D} does not find additional flux at larger radii. \citet{2014MNRAS.441.3417S} measure $\mathrm{m}_{814}^\mathrm{AB} = 18.54$ from \hst, which is only 7\%\ different from $\mathrm{m}_{814}^\mathrm{AB} = 18.61$ inferred from the ground-based Legacy data\footnote{To estimate the Legacy flux in the \hst\ passband, we fit an SED model to the Legacy Survey data and process it through {\texttt stsynphot}. These fluxes are not corrected for Galactic extinction.}. Radio emission tracing \qty{>200}{}\,\smpy\ star formation is unresolved with \qty{2}{\arcsecond} resolution imaging \citep{2020ApJ...901..138P}. Line emission tracing stellar photoionization is confined primarily to the inner $\sim$1\arcsec\ \citep{2023ApJ...947...33R}.

At intermediate radii of \qtyrange{1}{2}{\arcsecond}, the UV mission seen in the SBC data is negligibly contaminated by the wings of the nuclear point spread function (PSF). By finely sampling the F150LP PSF, \citet{2016ApJ...828...49H} find that the surface brightness of a point source declines to \num{\approx 5e-5} times the peak at $r = \qty{2}{\arcsecond}$. At a coarser radial binning of \qty{0.1}{\arcsecond}---more like the binned Makani data---this contrast decreases to \num{\approx 1e-4} \citep{2016acs..rept....5A}. In Makani, the surface brightness at the same radius from the peak is \numrange{\approx 5e-3}{1e-2} times the peak brightness (Figure~\ref{fig:rad-prof-cts}(a))---i.e., at least 50$\times$ the expected contribution from the PSF.

Nebular continuum from excited and/or ionized hydrogen is another possible emission channel in the UV-optical. \ion{H}{2} regions produce free-bound and free-free emission that dominates in the optical, while 2-photon (2$\gamma$) emission from 2s$\rightarrow$1s peaks at \qty{2000}{\text{\AA}} \citep{2024arXiv240803189K}. At the wavelengths of concern, only the 2$\gamma$ emission could contribute, and only at wavelengths \qty{>1216}{\text{\AA}}. \citet{2009ApJ...690...82D} show that this emission can, in fact, arise in cooling gas (see Section~\ref{sec:cooling}). However, the predicted equivalent width of \lya\ in such cooling gas, when compared to the nebular continuum, is \qty{>1000}{\text{\AA}} \citep{2009ApJ...690...82D}. Given that only \qty{\approx150}{\text{\AA}} of continuum lies within the filter above \lya\ (Figure~\ref{fig:throughputs}), this contribution might account for $\sim$15\%\ of any photons attributed to \lya. We ignore this contribution when calculating any $\lya$ properties. Regardless, the procedure below removes this contribution from the F150LP light along with \lya.

Reflection of 1350--2800~\AA\ emission from dust in the inner halos of galaxies and in galactic winds is observed in nearby star-forming galaxies \citep{2005ApJ...619L..99H, 2016ApJ...833...58H}. This process could contribute to the F150LP and F165LP bands at radii out to 15--20~\unit{\kpc} (\qty{3}{\arcsecond}), as the scattering albedo is high down to these rest wavelengths \citep{2003ApJ...598.1017D}. Dust extinction and continuum emission are observed out to these radii in Makani, as are neutral atomic and molecular phases that trace dusty gas (\citealt{2019Natur.574..643R, 2023ApJ...947...33R}; Veilleux et al., in preparation). We cannot at present directly constrain the magnitude of possible UV scattered light, though the UV emission does not obviously correlate morphologically with the dust emission (Veilleux et al., in preparation). In the analysis below we assume this contribution is negligible, but return to the possibility when it is salient.

\begin{figure}[t!]
\includegraphics[width=\columnwidth]{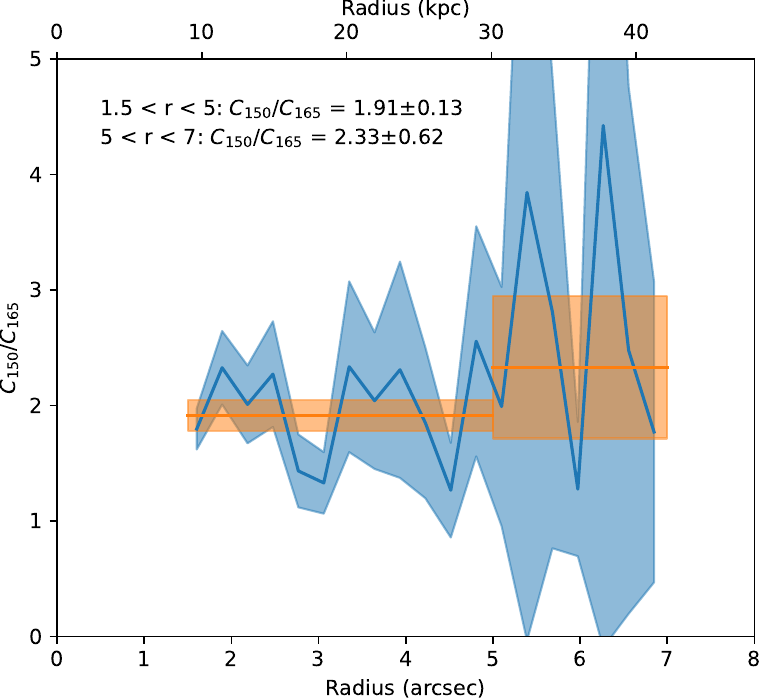}
\caption{The ratio $\mathcal{R} \equiv C_{150}/C_{165}$ vs. radius outside the continuum-dominated regions and out to the nebula edge. The blue line and shaded region indicate a radial profile and 1$\sigma$ error. The orange lines and shaded regions indicate the mean and standard deviation in the two circular annuli shown in the legend.}
\label{fig:c150c165-vs-rad}
\end{figure}

Thus, at radii $r \ga \qty{1.5}{\arcsecond}$, the emission is dominated by line emission rather than continuum emission, aside from a possible contribution from starlight scattered by dust at $\qty{1.5}{\arcsecond} < r < \qty{3}{\arcsecond}$. Together with \osix\ and \lya, \citet{2016ApJ...828...49H} consider other emission lines that potentially arise in the narrow bandpasses of these filters: \lyb, scattered \ion{C}{2} \qty{1036}{\text{\AA}}, and fluorescent \ion{C}{2}$^*$ \qty{1036}{\text{\AA}}. Based on the lack of observed \lyb\ or \ion{C}{2} in local starbursts \citep{2009ApJS..181..272G, 2011ApJ...730....5H, 2015ApJ...809...19H, 2016ApJ...828...49H}, the contribution of these lines is likely negligible. \lya\ halos in these systems arise due to many resonant scatterings, while \lyb\ is subject to re-emission as H$\alpha$ photons. Even if \lyb\ arose from direct emission in shocks, it is likely much fainter than \lya\ \citep{2020A&A...643A.101L}. {A high-ionization line that often accompanies \osix\ in quasar spectra is \ion{N}{5}~1238,~1243~\AA\ \citep[e.g.,][]{2022ApJ...926...60V}; this could appear in the red tail of the F165LP filter. However, it is expected to be weak compared to \osix\ in shocked gas \citep{2008ApJS..178...20A}, and is much less luminous than \lya\ even in active galactic nuclei.} These considerations imply that the extended emission in F150LP is due solely to \osix\ and \lya, while in F165LP it is due only to \lya.

The presence of both \osix\ and \lya\ in F150LP is indicated by the higher count rate and larger observed extent in F150LP compared to F165LP (Section~\ref{sec:morph}). To quantify the relative contribution, we use the count rate ratio $\mathcal{R}\equiv C_{150}/C_{165}$. Due to the low $S/N$ in the F165LP image (Figure~\ref{fig:img2}(e)), particularly at $r > \qty{5}{\arcsecond}$ (Figure~\ref{fig:rad-prof-cts}(b)), it is difficult to compute $\mathcal{R}$ across the nebula, even by Voronoi bin or radial profile. Instead, we compute two values, one each for the inner and outer regions of the nebula. In Figure~\ref{fig:c150c165-vs-rad} we show that $\mathcal{R}$ is fairly constant over the inner $\qty{1.5}{\arcsecond} < r < \qty{5}{\arcsecond}$ region in which F165LP shows significant emission, with $\mathcal{R}_\mathrm{inner} = 1.91\pm0.13$. In the outer, $\qty{5}{\arcsecond} < r < \qty{7}{\arcsecond}$ region, the ratio is much less constrained: $\mathcal{R}_\mathrm{outer} = 2.33\pm0.62$. This is consistent with a constant value of $\mathcal{R}$ but also with an increasing value with increasing radius. In other words, 44--51\%\ of counts in the F150LP image arise from \osix\ at $\qty{1.5}{\arcsecond} < r < \qty{5}{\arcsecond}$. At larger radius, the possible range is 42--66\%. 

\begin{figure*}[t!]
\includegraphics[width=\textwidth]{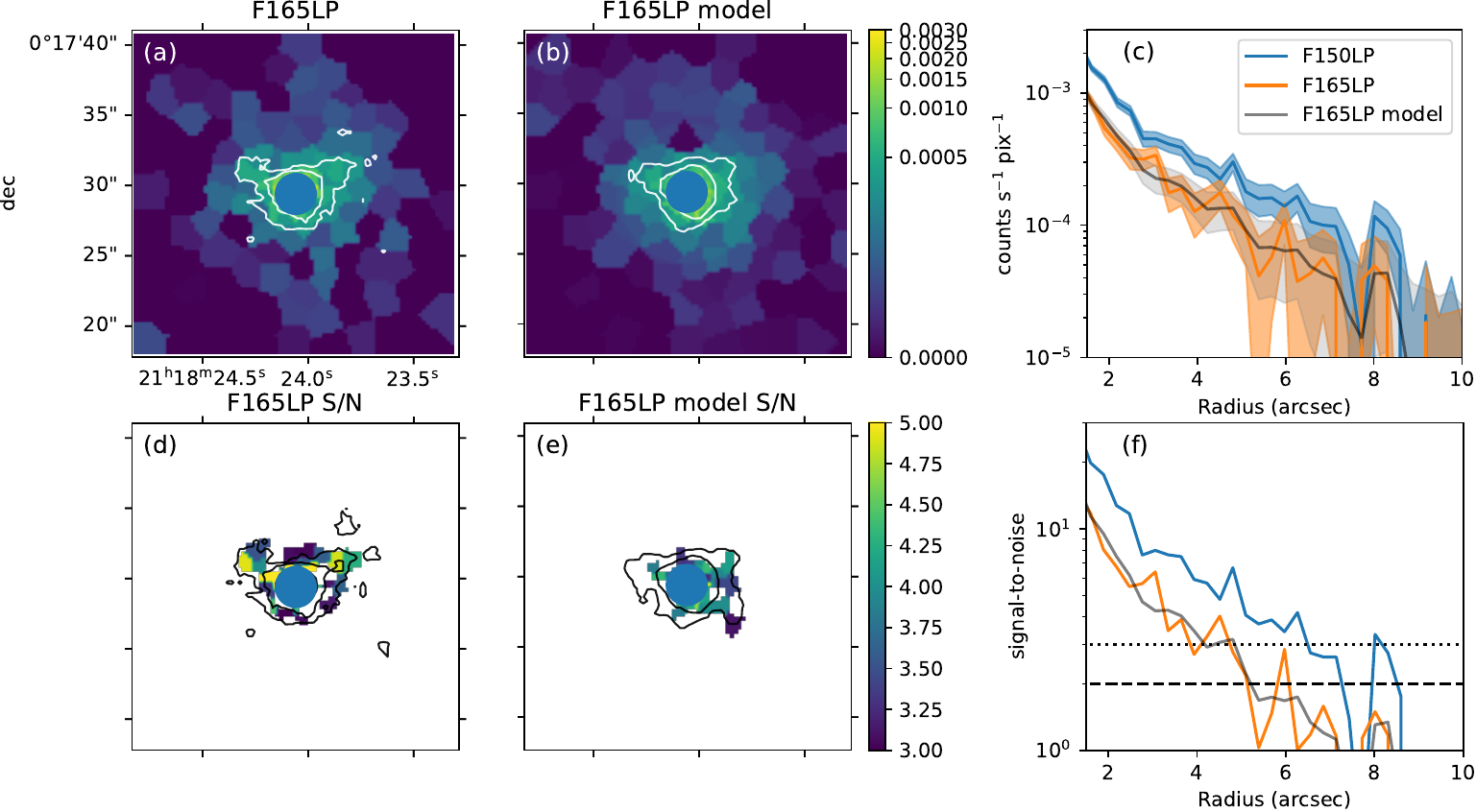}
\caption{Observed F165LP images and radial profiles, compared to those from a model using scaled F150LP data. (a) The Voronoi-binned and median-smoothed F165LP images of Makani, as in Figure~\ref{fig:img2}(b). We mask $r < \qty{1.5}{\arcsecond}$, which is dominated by continuum emission, and reduce the range of the colormap to emphasize lower fluxes. The contours are the same as in Figure~\ref{fig:img2}(b). (b) A model F165LP image, created by scaling the F150LP image assuming $C_{165} = \mathcal{R}_\mathrm{inner} C_{150}$ over $\qty{1.5}{\arcsecond} < r < \qty{5}{\arcsecond}$ and $C_{165} = \mathcal{R}_\mathrm{outer} C_{150}$ at $r > \qty{5}{\arcsecond}$. Note the similarity to panel (a). (c) Radial profiles of the observed F150LP and F165LP images and the model F165LP image, zooming in on radii expected to be emission-line dominated. The model and observed F165LP profiles are similar, by construction. (d) Same as Figure~\ref{fig:img2}(d), but modified as described for panel (a). Only bins with $S/N>3$ are shown. (e) A model F165LP S/N map, where the signal is from the model F165LP image, while the noise is equal to the observed F165LP values. Note the similar morphology to panel (d), illustrating that the apparent compactness of the F165LP image may be driven by the larger noise compared to F150LP, coupled with a lower signal. (f) Radial S/N profiles. Again, this illustrates that the apparent differences in morphology between F150LP and F165LP may be due to increased noise.}
\label{fig:f165-mod}
\end{figure*}

The lack of extended emission in the difference image $\mathrm{F150LP}-\mathrm{F165LP}$ (Figure~\ref{fig:img2}(c) and (f)) appears to contradict this. I.e., at face value it implies there is no \osix\ emission in the extended nebula. Subtracting F165LP---which contains only \lya---removes any significant signal. However, as we discuss in Section~\ref{sec:morph}, this is driven by Poisson statistics from dark noise in the F165LP image. To demonstrate this we create a model of the F165LP image from the F150LP data. We scale F150LP by the measured values of $\mathcal{R}_\mathrm{inner}$ and $\mathcal{R}_\mathrm{outer}$ over the corresponding radii to mimic the F165LP signal but keep the measured noise values from the F165LP data. The resulting model signal strongly resembles the observed F165LP image and radial profile of both signal and S/N (Figure~\ref{fig:f165-mod}). We thus conclude that the small apparent size of F165LP, and even smaller size of the difference image, is due to the increased noise and decreased signal when the images are subtracted, rather than a deficit of \osix\ emission in the F150LP image.

\subsection{Line Fluxes and Line Ratios} \label{sec:fluxes}

To convert count rates to astrophysical fluxes, we use the throughput $\mathcal{T}$ in F150LP and F165LP at the wavelengths of redshifted \osix\ and \lya. In the emission-line only region, the count rates in each filter are given by
\begin{flalign}
    && C_{150} = &\,\mathcal{C}\left(F_\mathrm{OVI}\mathcal{T}_\mathrm{OVI}^{150}+F_\mathrm{\lya}\mathcal{T}_\mathrm{\lya}^{150}\right) &&  \label{eq:c150}\\
    \mathrm{and} && C_{165} = &\,\mathcal{C}\left(F_\mathrm{\lya}\mathcal{T}_\mathrm{\lya}^{165}\right), &&  \label{eq:c165}
\end{flalign}
where $\mathcal{C}_j$ is the conversion from flux to counts in filter $j$, $F_i$ are the intrinsic fluxes of each line $i$, and $\mathcal{T}_i^j$ are the throughput values at the wavelength of the given line $i$ and filter $j$. Here we assume the wavelength dependence of the ACS-SBC response is captured completely in $\mathcal{T}$ and $\mathcal{C}$ is thus a constant.

\begin{figure}[t!]
\includegraphics[width=\columnwidth]{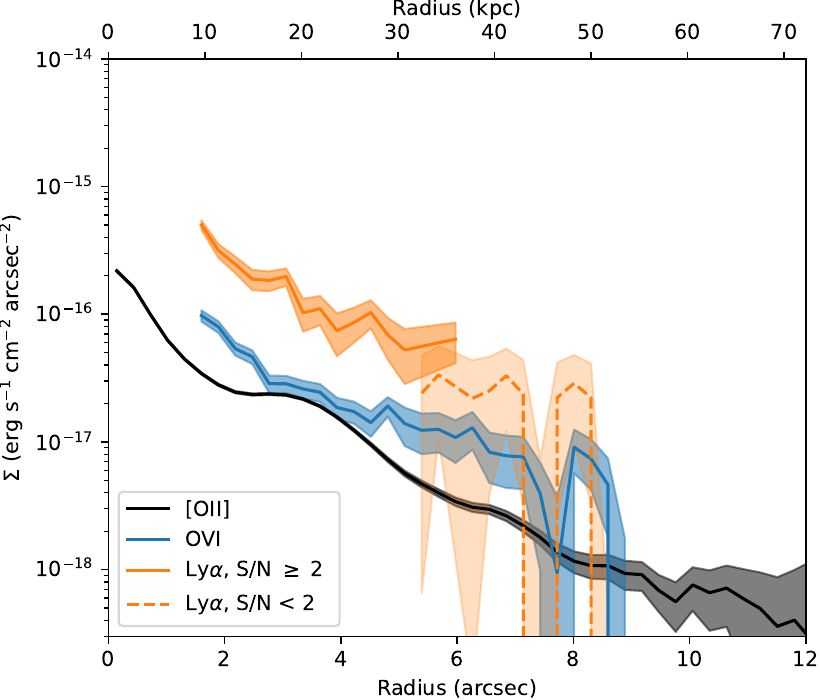}
\caption{Radial surface brightness profiles of \osix\ and \lya, with \ot\ \citep{2019Natur.574..643R} shown for comparison. We show only radial bins dominated by line emission. We have converted counts to line fluxes in F150LP and F165LP using Eq.~\ref{eq:fovinum} and \ref{eq:flya} and flux to surface brightness using the drizzled pixel scale (Section~\ref{sec:red}). We plot the high-significance and low-significance \lya\ bins separately. Shaded regions represent 1$\sigma$ errors, including uncertainties in both count rates ($C_{150}$ and $C_{165}$) and average $\mathcal{R}$ (Figure~\ref{fig:c150c165-vs-rad}).}
\label{fig:rad-prof-fl}
\end{figure}
Combining Equations~\ref{eq:c150} and \ref{eq:c165}, we express the \osix\ flux as
\begin{equation} \label{eq:fovi}
F_\mathrm{OVI} = \, \frac{C_{150}}{\mathcal{C}\mathcal{T}_\mathrm{OVI}^{150}}\left(1-\frac{\mathcal{T}_\mathrm{\lya}^{150}}{\mathcal{T}_\mathrm{\lya}^{165}}\mathcal{R}^{-1}\right).\\
\end{equation}
(As a reminder, we defined $\mathcal{R}= C_{150}/C_{165}$ in Section~\ref{sec:source}.) We assume that the F150LP image---outside of $r \ga \qty{1.5}{\arcsecond}$---describes the morphology of both \osix\ and \lya. We estimate the quantity $\mathcal{C}^{150}_\mathrm{OVI}\equiv \mathcal{C}\mathcal{T}_\mathrm{OVI}^{150}$ using a synthetic emission line profile input to \texttt{synphot} and \texttt{stsynphot} \citep{2018ascl.soft11001S, 2020ascl.soft10003S}, while $\mathcal{T}_\mathrm{\lya}^{150}$ and ${\mathcal{T}_\mathrm{\lya}^{165}}$ come directly from the response function (Section~\ref{sec:obs}, Figure~\ref{fig:throughputs}). The result is 
\begin{equation}
F_\mathrm{OVI} =\, (\qty{7.68e-15}{\erg\per\second\per\cm\squared})C_{150}(1-1.07\mathcal{R}^{-1}) \label{eq:fovinum}
\end{equation}
where $C_{150}$ is in \unit{\count\per\second}. 

\begin{figure*}[ht!]
\includegraphics[width=\textwidth]{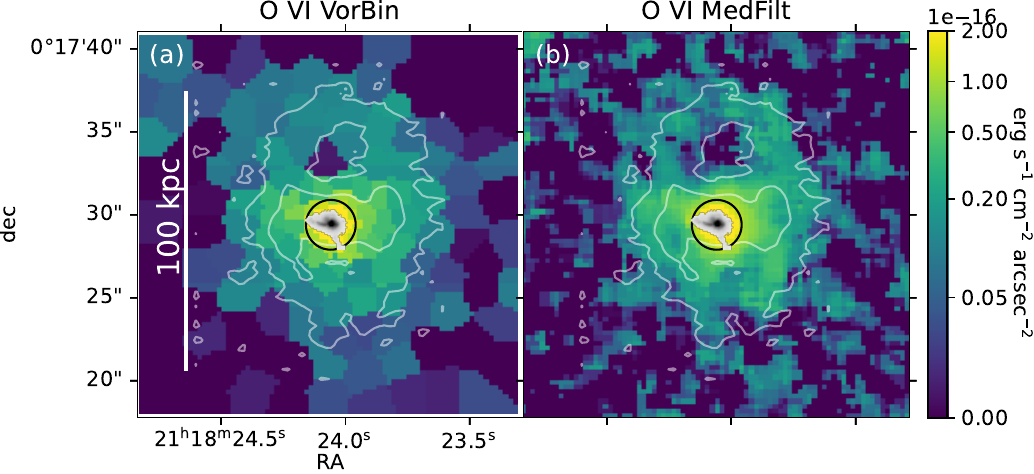}
\caption{Voronoi-binned and median-filtered flux maps of \osix, as computed from Equation~\ref{eq:fovinum}. We use the measured values of $\mathcal{R}_\mathrm{inner}$ and $\mathcal{R}_\mathrm{outer}$, shown in Figure~\ref{fig:c150c165-vs-rad}, to correct for $\lya$ contamination. We delineate the inner, continuum-dominated region with a \qty{1.5}{\arcsecond} circle and overlay the F814W starlight image.}
\label{fig:fluxed-o6}
\end{figure*}

To compute $F_\mathrm{\lya}$, we can invert Equation~\ref{eq:c165} and input $C_{165}$. As before, we use \texttt{synphot} and \texttt{stsynphot} to estimate $\mathcal{C}^{165}_\mathrm{\lya}\equiv \mathcal{C}\mathcal{T}_\mathrm{\lya}^{165}$. Alternatively, we can use the definition of $\mathcal{R}$ and rearrange Equation~\ref{eq:c165} to express the \lya\ flux in terms of $C_{150}$:
\begin{equation} \label{eq:flya}
F_\mathrm{\lya} = \, \frac{C_{150}}{\mathcal{C}\mathcal{T}_\mathrm{\lya}^{165}}\mathcal{R}^{-1}= \, (\qty{3.20e-14}{\erg\per\second\per\cm\squared})C_{150}\mathcal{R}^{-1} \\
\end{equation}
These yield similar results. For a more conservative approach, we adopt the simple inverse of Equation~\ref{eq:c165}.
We apply Equation~\ref{eq:fovinum} to the radial profile of $C_{150}$ and Equation~\ref{eq:flya} to the radial profile of $C_{165}$ (Figure~\ref{fig:rad-prof-cts}(a)) over $\qty{1.5}{\arcsecond} < r < \qty{9}{\arcsecond}$. This ignores the inner region contaminated by continuum but extends outward to the 2$\sigma$ detection region of the $C_{150}$ map (Figure~\ref{fig:rad-prof-cts}(b)). We use the measured values of $\mathcal{R}_\mathrm{inner}$ and $\mathcal{R}_\mathrm{outer}$, with the latter applying at all radii $r > \qty{5}{\arcsecond}$. We also correct for Milky Way extinction using reddening maps derived from Dark Energy Spectroscopic Instrument (DESI) Survey observations of stellar spectra. We take the average of the ($g-r$) and ($r-z$) maps, $\ebv=0.0587$ \citep{2024arXiv240905140Z}. We apply the Milky Way extinction model with $R_\mathrm{V} = 3.1$ from \citet{2023ApJ...950...86G} using the \texttt{dust\_extinction} package \citep{Gordon2024}. The resulting radial profiles are similar in shape to that of \ot\ (Figure~\ref{fig:rad-prof-fl}). We apply the same procedure to produce fluxed maps of \osix\ emission (Figure~\ref{fig:fluxed-o6}).

\begin{figure*}[ht!]
\includegraphics[width=\textwidth]{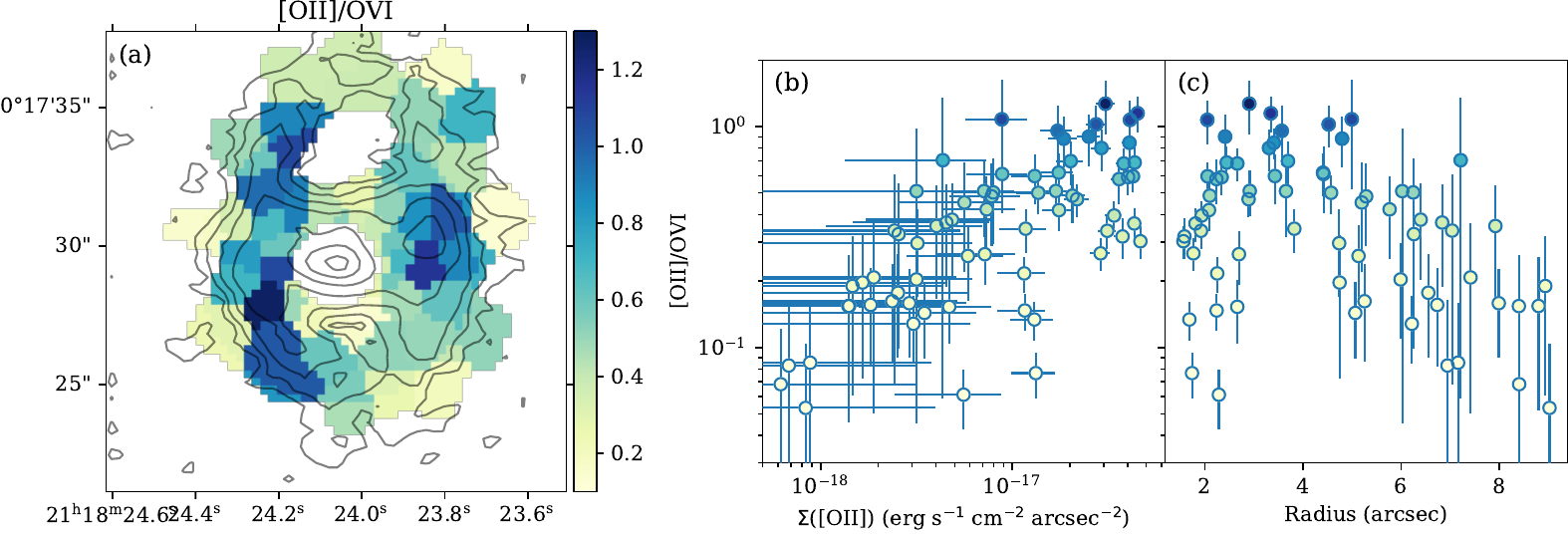}
\caption{(a) Map of flux ratio $F(\text{\otl})/F(\text{\osixl})$, with \ot\ contours overlaid. We mask the $r < \qty{1.5}{\arcsecond}$ region, and show only bins with $\mathrm{S/N} > 1$ in both lines. The ratio appears highest East and West, while it is lower directly North and South. (b)--(c): Line ratio vs. \ot\ surface brightness and radius in each Voronoi bin. The errors shown are 1$\sigma$, and the bins are colored by their line ratio as in (a). \otl/\osix\ increases with increasing $\Sigma(\ot)$. It peaks at radii \qtyrange{2}{5}{\arcsecond}, primarily along the East and West sides of the hourglass shape, and declines at lower and higher radii.}
\label{fig:o2o6}
\end{figure*}

With our flux-calibrated \osix\ image, we can compute the observed (attenuated) line ratio \ot/\osix\ in each Voronoi bin, which we show in Figure~\ref{fig:o2o6}. The resulting values are in the range 0.05--1.3. They appear to be elevated to the East and West of the nucleus along the rims of the hourglass shape, and lower directly North and South. We also plot these values against the \ot\ surface brightness $\Sigma$(\ot) and projected radius. It is apparent that higher \ot/\osix\ values correspond to regions of higher \ot\ surface brightness and decline at both small and large radius.
%Figure~\ref{fig:o2o6}(b) shows that this is in part a radial effect: at any value of $\Sigma$(\ot), larger values of \ot/\osix\ are found at larger radii. However, at fixed radius, there is still an increase of \ot/\osix\ with $\Sigma$(\ot).

Combining Equations~\ref{eq:c150} and \ref{eq:c165}, we can also write the flux ratio of the two lines in the extended nebula as
\begin{equation} \label{eq:ovilya}
    \frac{F_{OVI}}{F_\mathrm{\lya}} =\, \frac{\mathcal{T}_\mathrm{\lya}^{150}}{\mathcal{T}_{OVI}^{150}}\left(\mathcal{R}\frac{\mathcal{T}_\mathrm{\lya}^{165}}{\mathcal{T}_{\lya}^{150}} - 1\right) = \, 0.20(0.93\mathcal{R} - 1)
\end{equation}
Using our measured values for $\mathcal{R}$, and correcting for Milky Way extinction as above, we thus find $(F_{OVI}/F_\mathrm{\lya})_\mathrm{inner}=0.16\pm0.03$ and $(F_{OVI}/F_\mathrm{\lya})_\mathrm{outer}=0.24\pm0.12$. 

\section{Discussion} \label{sec:discuss}

We have shown that the Makani wind nebula, previously imaged in the low ionization emission lines of \ot\  \citep{2019Natur.574..643R}, also shows high ionization \osix\ and \lya\ emission that is spatially coincident with \ot\ throughout the observed volume.

\subsection{Integrated Wind Properties} \label{sec:wind-prop}

As we discuss in Section~\ref{sec:intro}, the single, spatially-resolved image of an \osix\ halo is J1156 \citep{2016ApJ...828...49H}. This $z = 0.235$ galaxy is a low-mass ($10^9$~\msun) galaxy with star formation rate $\mathrm{SFR}\approx 30$~\smpy. The radial surface brightness profile of this galaxy is consistent with an exponential profile with central surface brightness $\Sigma_0 = \qty{5.3e-17}{\erg\per\second\per\cm\squared\per\arcsec\squared}$ and exponential radius $r_\mathrm{exp} = \qty{7.5}{\kpc}$. This is equivalent to an $n=1$ S\'{e}rsic with effective (half-light) radius $r_\mathrm{e} = 1.6783 \ r_\mathrm{exp} = \qty{12.6}{\kpc}$, with an extrapolated surface brightness at this radius of $\Sigma_\mathrm{e} = \qty{2.0e-17}{\erg\per\second\per\cm\squared\per\arcsec\squared}$. The scale of the \osix\ halo is about 10$\times$ the scale of the photoionized, \ha- and \lya-emitting gas in this system \citep{2016ApJ...828...49H}. \citet{2016ApJ...828...49H} infer a density of $n_\mathrm{e} \approx \qty{0.5}{\per\cm\cubed}$ for the coronal gas, and its kinematics imply an outflow.
%They conclude that this gas arises at the interface of the hot, outflowing gas with cool clouds in the wind.

\begin{figure}[t!]
\includegraphics[width=\columnwidth]{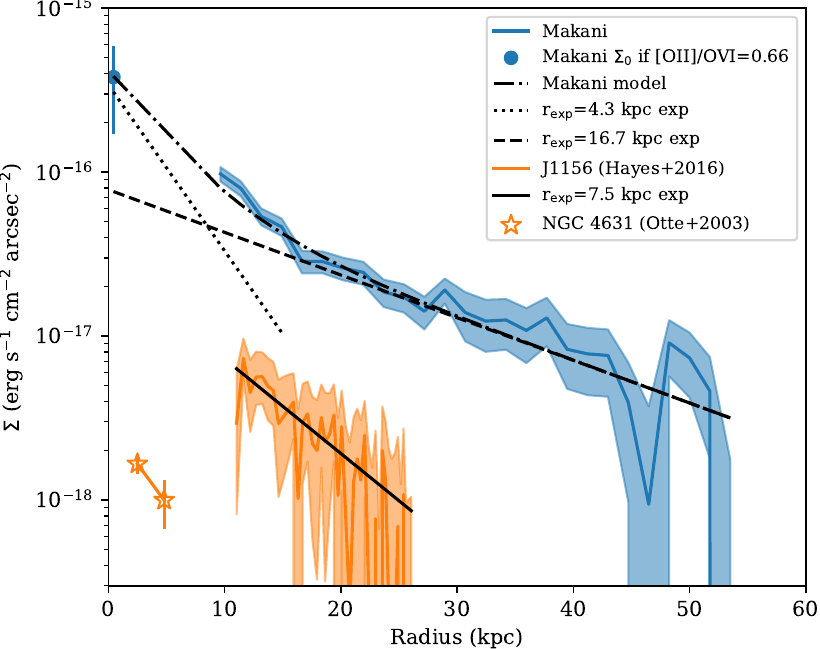}
\caption{Surface brightness profile of \osix\ emission in Makani, J1156 \citep{2016ApJ...828...49H}, and NGC~4631 \citep{2003ApJ...591..821O} as a function of projected radius. The nuclear point of Makani is derived from \ot. A double exponential model (Eq.~\ref{eq:twoexp}) is overplotted. The J1156 and NGC~4631 data have been scaled downward by factors of \qty{\approx2}{} and \qty{\approx5}{}, as if they were observed at the redshift of Makani. The best-fit exponential model from \citet{2016ApJ...828...49H} is also shown. The Makani nebula is significantly brighter and larger than that of J1156---with 20$\times$ the total estimated luminosity---consistent with its 10$\times$ higher star-formation rate. The two-component fit may reflect the two outflow episodes. The inner exponential, containing 20\%\ of the \osix\ light, may also have some contribution from dust scattering of starlight (Section~\ref{sec:source}).}
\label{fig:sbcomp}
\end{figure}
    
In Figure~\ref{fig:sbcomp} we compare J1156 to the radial surface brightness profile of Makani. As a reference point, we estimate the central surface brightness $\Sigma_0$(\osix) by scaling $\Sigma_0$(\ot) using $\ot/\osix\ = 0.66\pm0.29$. This is the mean and standard deviation measured in the Voronoi bins with $\Sigma(\ot) > \qty{2e-17}{\erg\per\second\per\cm\squared\per\arcsec\squared}$ (Figure~\ref{fig:o2o6}(b)). We multiply the J1156 surface brightness by 0.51, which is the relative $(1+z)^4$ dimming of itself and Makani, as if J1156 were observed to be at $z = 0.459$ rather than 0.235. We compare also in this figure to the surface brightness of the \osix\ detection in NGC~4631, which is of order \qty{1e-18}{\erg\per\second\per\cm\squared\per\arcsec\squared} at \qty{4}{\kpc} radius, if it were observed at the redshift of Makani \citep{2003ApJ...591..821O}.

The profile of \osix\ emission in Makani is well-fit by a double exponential profile. We use the \texttt{emcee} \citep{2013PASP..125..306F} minimizer in \texttt{lmfit} \citep{2021zndo....598352N} to find the best-fit double exponential:
\begin{equation} \label{eq:twoexp}
    \Sigma(r) = \Sigma^\mathrm{in}_0 e^{-r/r_\mathrm{exp}^\mathrm{in}} + \Sigma^\mathrm{out}_0 e^{-r/r_\mathrm{exp}^\mathrm{out}}.
\end{equation}
We set the upper bound of $r_\mathrm{exp}^\mathrm{in}$ equal to the lower bound of $r_\mathrm{exp}^\mathrm{out}$. We also fix the amplitude $\Sigma^\mathrm{in}_0$ of the inner profile to equal 90\%\ of the total amplitude $\Sigma_0$, as estimated from \ot. In Table~\ref{tab:ovi} we list the best-fit parameters.
\begin{deluxetable}{CC}
    \label{tab:ovi}
    \tablecaption{\osix\ Measurements}
    \tablewidth{0pt}
    \tablehead{\colhead{Quantity} & \colhead{Value}}
%    \colnumbers
%    \decimals
    \startdata
    \multicolumn{2}{c}{Best-fit 2-exponential model parameters} \\
    \hline
\Sigma^\mathrm{in}_0 & 3.42\times10^{-16}~\unit{\erg\per\second\per\cm\squared\per\arcsec\squared} [fixed]\\
r_\mathrm{exp}^\mathrm{in} & (4.29\pm0.34)~\unit{\kpc} \\
\Sigma^\mathrm{out}_0 & (0.78\pm0.10)\times10^{-16}~\unit{\erg\per\second\per\cm\squared\per\arcsec\squared}\\
r_\mathrm{exp}^\mathrm{out} & (16.68\pm1.01)~\unit{\kpc} \\
    \hline
    \multicolumn{2}{c}{Voronoi binned data, $S/N > 2, r > \qty{1.5}{\arcsec}$} \\
    \hline
\mathrm{Flux} & (2.89\pm0.10)\times10^{-15}~\unit{\erg\per\second\per\cm\squared} \\
\mathrm{Luminosity} & (2.42\pm0.09)\times10^{42}~\unit{\erg\per\second} \\
    \hline
    \multicolumn{2}{c}{2-exponential model, extrapolated to $\infty$} \\
    \hline
\mathrm{Flux} & 4.86\times10^{-15}~\unit{\erg\per\second\per\cm\squared} \\
\mathrm{Luminosity} & 4.06\times10^{42}~\unit{\erg\per\second} \\
r_\mathrm{eff} & 21.42~\unit{\kpc} \\
    \hline
    \multicolumn{2}{c}{2-exponential model, extrapolated to 60~\unit{\kpc}} \\
    \hline
\mathrm{Flux} & 4.39\times10^{-15}~\unit{\erg\per\second\per\cm\squared} \\
\mathrm{Luminosity} & 3.67\times10^{42}~\unit{\erg\per\second} \\
r_\mathrm{eff} & 18.85~\unit{\kpc} \\
    \enddata
%    \tablecomments{}
\end{deluxetable}

To compute the integrated properties of the nebula, we take several approaches. First, we integrate the flux of the nebula over the observed Voronoi bins with $r > \qty{1.5} {\arcsec}$ and $S/N > 2$ detections. Second, we extrapolate the double exponential profile to \qty{60}{\kpc} and to $\infty$, for comparison. The first method does not include the unmeasured nuclear flux and yields about three-quarters of the flux of the profile integration method. Integrating to $\infty$ rather than \qty{60}{\kpc} increases the flux by about 10\%. 22\%\ of the total flux resides under the inner exponential. We find an effective (half-light) radius of \qtyrange{19}{21}{\kpc}, similar to that of \ot, for which $r_\mathrm{eff} \approx \qty{18}{\kpc}$ \citep{2019Natur.574..643R}. We also list these values in Table~\ref{tab:ovi}.

The Makani nebula is produced by multiple outflow episodes over several hundred Myr. It may thus be that the inner exponential corresponds to the recent, Episode II emission, while the outer exponential represents the more evolved, Episode I nebula. The entire \osix\ nebula is 20$\times$ more luminous than that of J1156. This higher luminosity is consistent with the increase of CGM \osix\ absorbers with star formation rate \citep{2023ApJ...949...41T}. Makani's SFR is 224--300~\smpy\ \citep{2020ApJ...901..138P}, vs. 30~\smpy\ in J1156. Similarly, the \osix\ surface brightness of NGC~4631---with SFR of 3~\smpy\ \citep{2015ApJ...804...46M}---is of order 10$\times$ smaller even than J1156 at comparable radii. These scalings are broadly consistent with the roughly linear relationship of $L_\mathrm{OVI}$ and SFR surface density in the simulations of \citet{2017ApJ...835L..10L}. Alternatively, some of the inner exponential in Makani could arise from starlight scattered off dust grains (Section~\ref{sec:source}). This would amount to at most 20\%\ of the total \osix\ flux if the entirety of the inner exponential was due to this process.

The \lya\ nebula in Makani is also bright and large. We apply Equation~\ref{eq:flya} to the $C_{165}$ map and correct for Milky Way extinction. As with \osix, we then integrate over Voronoi bins with $r > \qty{1.5} {\arcsec}$ and $S/N > 2$ detections. {The resulting flux and luminosity are $(8.7\pm0.5)\times10^{-15}$~\unit{\erg\per\second\per\cm\squared} and $(7.3\pm0.4)\times10^{42}$~\unit{\erg\per\second}. If the half-light radius of \lya\ is similar to that of \osix, this places the Makani nebula in the \lya\ blob class \citep{2020ARA&A..58..617O}. In this context, it is consistent with other \lya\ blobs that arise around massive, star-forming galaxies and AGN.

%We do not have a direct measure of the inner \lya\ surface brightness, though if we scale the H$\alpha$ surface brightness from \citet{2023ApJ...947...33R} by the expected $\lya/\ha$ in the absence of extinction \citep[see][and references therein]{2015PASA...32...27H}, an upper limit in the case of no \lya\ escape is $\sim$1.0}
%Using our inference that the \lya\ and \osix\ morphologies are similar, the measured \osix/\lya\ values, and $L$(\osix), then we compute $L(\mathrm{Ly}\alpha)\approx\qty{1e43}{\erg\per\second}$. 

\subsection{In-situ Emission vs. Scattered Light} \label{sec:scattering}

\osix\ in the CGM has to date been observed primarily in absorption against background sightlines. Like other transitions that lead to absorption in diffuse interstellar gas, the lower level of the \osixl\ transitions is the ground state \citep{1991ApJS...77..119M}. \osix\ emission is produced by in-situ processes that excite the O$^{+5}$ ions, like collisional heating or irradiation by local sources. However, as a so-called resonant line, it is also a channel for continuum scattering into the line of sight. Sighting ``down the barrel" toward the starlight from a galaxy, an outflowing shell absorbs continuum at blueshifted wavelengths and scatters photons back to the observer at redshifted wavelengths. This is the classic P-Cygni profile. More complicated gas and dust distributions result in other lineshapes for resonant lines in a galactic wind, which are mainly variations on this theme \citep{2011ApJ...734...24P}.

A spherically-symmetric, expanding nebula with no dust is the simplest galactic wind. When the emerging light is spatially and spectrally integrated over the continuum background source and larger-scale wind, the total equivalent width of a resonant line is zero \citep{2006A&A...460..397V, 2011ApJ...734...24P}. That is, the emission and absorption equivalent widths from scattering are equal in magnitude. Using this simple model, \citet{2016ApJ...828...49H} estimated at least 6$\times$ more \osix\ emission in J1156 overall than that observed only along the line of sight aperture to the starlight. In the starlight-only aperture covering J1156, the emission and absorption equivalent widths are approximately equal. Thus, the scattered light in this approximation accounts for at most 1/6th of the observed \osix\ emission in J1156. In Haro~11, another starburst with similar star formation rate to J1156, the integrated absorption and emission equivalent widths are approximately equal \citep{2007ApJ...668..891G}.

We have no \osix\ spectroscopy of Makani from which to construct a similar argument. However, we can assume the spherically-symmetric model and consider the possible magnitudes of line emission and absorption along the line of sight to the galaxy starlight. If the extended \osix\ emission we measure is stronger than the possible backscattered emission along the same line of sight as any possible absorption---i.e., along the line of sight to the galaxy's stars---than we can consider the emission to be dominated by in-situ emission.

The starlight footprint of Makani is highly concentrated, with an aperture of radius \qty{1}{\arcsecond} containing 80--90\%\ of the UV-optical continuum (Section~\ref{sec:morph}--\ref{sec:source}). Such an aperture contains 13\%\ of the \osix\ flux, as estimated from the 2-exponential model. To calculate the equivalent width of this emission, we assume the F150LP image is continuum-dominated at $r < \qty{1}{\arcsecond}$ and use PHOTFLAM to estimate the total flux density $F_\mathrm{1605}$ at the filter pivot wavelength. The resulting rest-frame equivalent width of emission is \qty{7}{\text{\AA}}.

For comparison, \citet{2009ApJS..181..272G} report 1032~\AA\ absorption-line equivalent widths of \qtyrange{0.2}{1.7}{\text{\AA}} in a sample of nearby starbursts. The total \osixl\ equivalent widths would then be 1.5--2$\times$ this value in the optically thin and thick limits, respectively, which yields a range of \qtyrange{0.3}{2.6}{\text{\AA}} or \qtyrange{0.4}{3.4}{\text{\AA}}. The measured full widths at half maximum of the \osix\ lines in this sample are \qtyrange{118}{763}{\km\per\second}. We verified that these linewidths correlate with the equivalent widths, much as they correlate with the column densities \citep{2002ApJ...577..691H,2009ApJS..181..272G,2017ApJ...848..122B}. The \ot\ linewidths in Makani average $\langle \mathrm{FWHM} \rangle = 900~\kms$ in the inner wind \citep{2023ApJ...947...33R}. Assuming that the \osix\ linewidths are at least as large, then we might expect an equivalent width of order \qtyrange{4}{5}{\text{\AA}}. Scaling from the linewidth, the corresponding absorption column density is $\mathrm{log}(N_\mathrm{OVI}/\mathrm{cm}^{-2}) \sim 15.3$, or 4$\times$ higher than in J1156 \citep{2016ApJ...828...49H}.

The similarity of these estimates of absorption and emission equivalent width towards the starlight suggests that any \osix\ emission interior to $\sim$\qty{1}{\arcsecond} is offset by corresponding absorption. The other $\sim$85\%\ of the \osix\ we observe in emission is then due to in-situ processes and not dominated by scattering. However, this simple scenario could be modified by dust that blocks backscattered \osix\ towards the nucleus and reduces redshifted emission, or alternatively that we observe the starlight along a low column-density sightline that reduces blueshifted absorption \citep{2011ApJ...734...24P}.

Resonant emission is observed in Makani in \nad\ and \ion{Mg}{2}, which are much lower ionization states \citep{2019Natur.574..643R, 2023ApJ...947...33R}. The \nad\ absorption and emission equivalent widths are approximately equal in magnitude, at a relatively small \qtyrange{0.4}{0.5}{\text{\AA}} in the central arcsecond \citep{2023ApJ...947...33R}. There is no detectable \ion{Mg}{2} absorption. However, both doublets show emission out to \qtyrange{10}{20}{\kpc}. If the emission from these lines are due to continuum scattering, then perhaps such scattering in these ions is not compensated for by line-of-sight absorption due to spherical asymmetry. However, this extended emission may also be line scattering from collisional excitation or photoionization \citep{2018ApJ...855...96H, 2020MNRAS.498.2554C}.

\subsection{Cool Clouds in a Hot Wind} \label{sec:cooling}

\begin{figure}[t!]
\fig{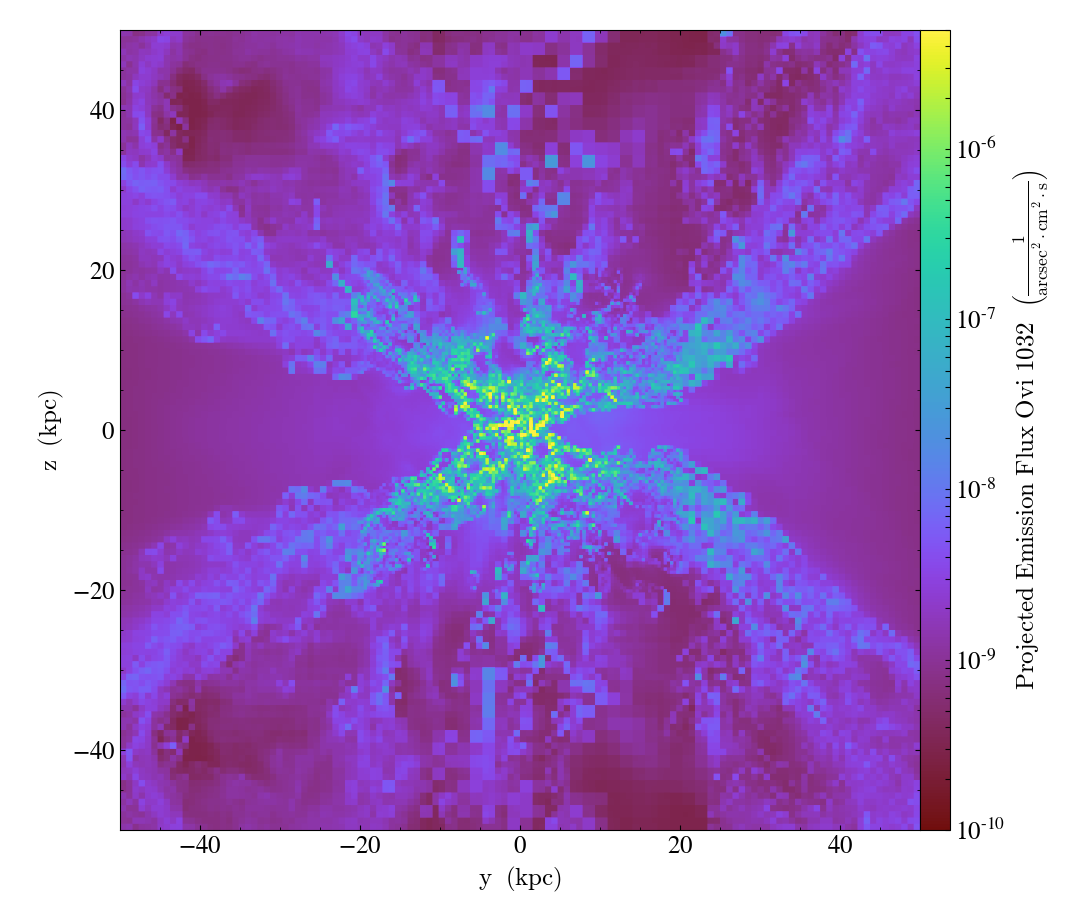}{0.49\textwidth}{(a)}
\fig{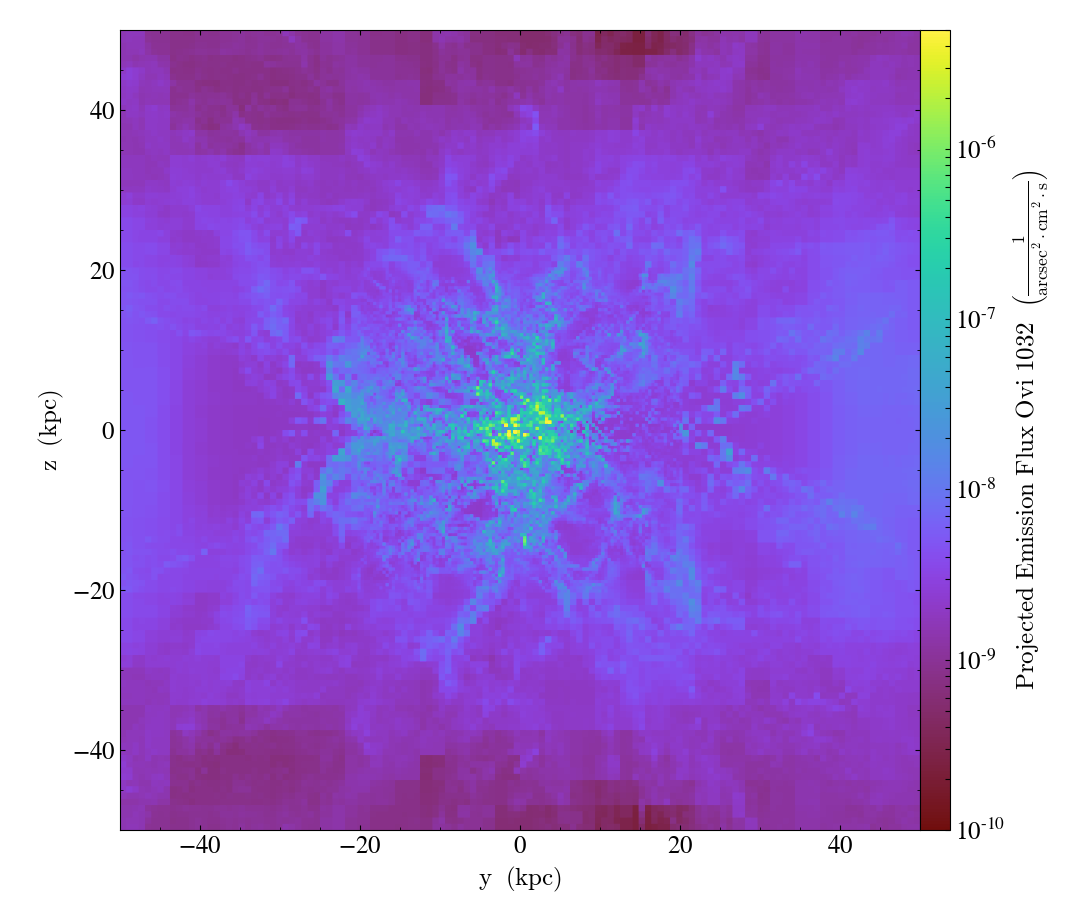}{0.49\textwidth}{(b)}
\caption{Projected \osix\ \qty{1032}{\text{\AA}} emission from the simulations of \citet{2020ApJ...898..148L}. Shown is a $6\times10^{10}$~\msun\ galaxy with a star formation rate of (a) 30~\smpy\ and (b) 200~\smpy. The units are \unit{\photon\per\second\per\cm\squared\per\arcsec\squared}. Assuming similar emission from both doublet lines, and reducing the simulated surface brightness to the redshift of Makani by $(1+z)^4$, $\qty{1e-6}{\photon\per\second\per\cm\squared\per\arcsec\squared} = \qty{1e-17}{\erg\per\second\per\cm\squared\per\arcsec\squared}$. The scale and brightness of the filamentary structure are similar to those observed in Makani (Figure~\ref{fig:fluxed-o6}). In this model, \osix\ arises primarily from the condensation of the hot CGM into cool clouds as a hot wind blows into the CGM.}
\label{fig:sim}
\end{figure}

This basic morphology implies that most of the \osix\ we observe arises not from the hot, volume-filling CGM, but rather from the interaction between cool, $T\approx10^4$~K clouds and the hot, $T>10^6$~K  haloes in which they are embedded \citep{2001ApJ...554.1021H,2018MNRAS.474.1688C,2022ApJ...924...82F}. This interaction causes hydrodynamic instabilities \citep[e.g.,][]{2015ApJ...805..158S}, which in turn produce turbulent mixing layers \citep[e.g.,][]{2018MNRAS.480L.111G} that emit high-ionization UV lines in the interface between the hot and cool gas \citep[e.g.][]{1993ApJ...407...83S}.  The clouds can grow as mass is transferred from the hot to the cool phase \citep{2018MNRAS.480L.111G, 2020ApJ...903...32F}. The gas cools radiatively, producing \osix\ as the temperature passes through $T\approx10^{5.5}$~K. The emissivity of \osix\ peaks at this temperature in collisional ionization equilibrium \citep{1993ApJS...88..253S}.

Conversely, mass transfer can also move in the other direction. The cold clouds are heated through, e.g., thermal conduction, and then mass-loaded into the hot, surrounding medium. This arises in the interface between the $10^7$~K wind fluid and entrained clouds in the launching region of a galactic wind. This leaves a signature in the X-ray emitting wind fluid \citep{2009ApJ...697.2030S}. It also produces velocity stratification, in that higher-velocity gas is more highly ionized as evaporation proceeds under cloud acceleration \citep{2001ApJ...554.1021H,2009ApJS..181..272G,2018MNRAS.474.1688C}.

As an example of how this emission may arise in large-scale winds as they propagate into the CGM, we show in Figure~\ref{fig:sim} the projected \osix\ emission from the simulations of \citet{2020ApJ...898..148L}. These simulations include mass loading into the hot wind from cloud evaporation at small scales \citep{2017ApJ...841..101L}. As the hot wind expands into the CGM, the hot halo gas ``saturates'' and cool clouds condense from the hot gas  in clumpy, filamentary structures. The simulations shown are for winds produced by a $6\times10^{10}$~\msun\ galaxy, which is 50--100\%\ of the mass of Makani \citep{2019Natur.574..643R, 2022AJ....164..222W}. The star formation rates of the two simulations are 30~\smpy and 200~\smpy, or 10\%\ and 70\%\ of the time-averaged SFR of Makani \citep{2020ApJ...901..138P}. The filamentary structure of \osix\ is similar to that observed in Makani (Figure~\ref{fig:fluxed-o6}), including the scale of the emission and the observed fluxes. In this model, the hot, outflowing wind is critical for the formation of the cool clouds observed in \ot, which also emit in \osix\ at the same physical location as the gas radiatively cools.

\subsection{Shock models} \label{sec:shocks}
Alternatively, the \osix\ emission could arise from shock ionization as the wind propagates through the CGM. In \citet{2023ApJ...947...33R} we argued that the warm ionized gas observed in strong, rest-frame optical lines is consistent with shock models. This conclusion was based a comparison of line ratios and kinematics to the predictions of \citet{2008ApJS..178...20A}. These models point to pre-shock densities in the range $n_\mathrm{H} = \qtyrange{1}{10}{\per\cm\cubed}$, with postshock densities rising to $n_\mathrm{H} = \qtyrange{600}{6000}{\per\cm\cubed}$. The larger densities apply to the inner ($r \la \qty{15}{kpc}$) Episode II wind with $\langle\sigma\rangle = \qty{400}{\km\per\second}$. The smaller apply to the outer, older Episode I wind with $\langle\sigma\rangle = \qty{200}{\km\per\second}$.

These same models predict fluxes for both \osix\ and \lya. In Section~\ref{sec:fluxes} we compute the flux ratios \osix/\lya\ and \ot/\osix. In Figure~\ref{fig:shock-models} we compare the observed ratios to model predictions. We show shock$+$precursor models with Solar and Large Magnellanic Cloud (LMC) abundances, and with densities $n = \qty{1}{\per\cm\cubed}$ and \qty{100}{\per\cm\cubed}, corresponding to models M, Q, and L of  \citet{2008ApJS..178...20A}. We vary the model grids over magnetic parameter $b\equiv Bn^{-1/2}$ and shock velocity.

\begin{figure*}[ht!]
\includegraphics[width=\textwidth]{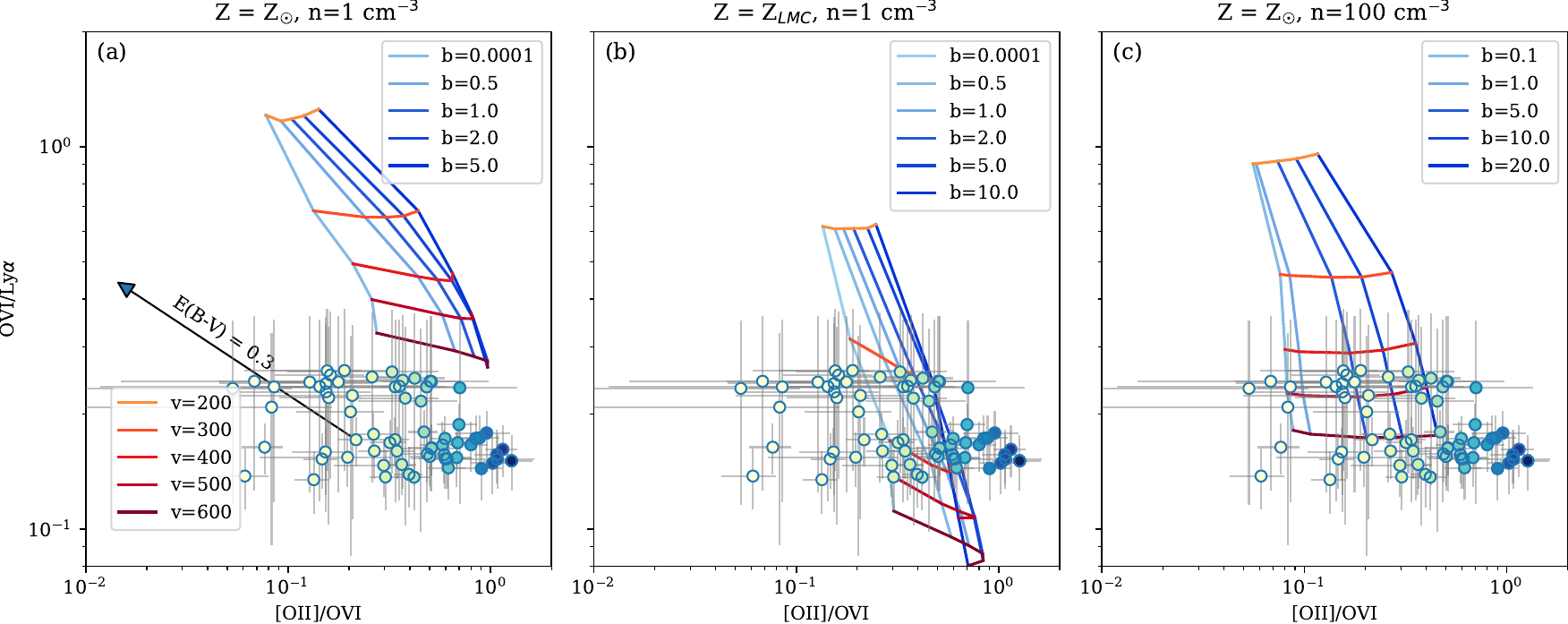}
\caption{Line ratios vs. shock model predictions. (a) Solar metallicity, $n=\qty{1}{\per\cubic\cm}$ grids \citep{2008ApJS..178...20A}. (b) Large Magellanic Cloud (LMC) metallicity, $n=\qty{1}{\per\cubic\cm}$ grids. (c) Solar metallicity, $n=\qty{100}{\per\cubic\cm}$ grids. \otl/\osixl\ ratios represent individual Voronoi bins (Figure~\ref{fig:o2o6}(a)). The \osix/\lya\ ratios for each point correspond to one of two average values calculated in radial bins (Figure~\ref{fig:c150c165-vs-rad} and Section~\ref{sec:fluxes}); we have also added a small vertical random scatter to each point for display purposes. Error bars are $1\sigma$. With the arrow, we show the effect of $\ebv=0.3$ mag, which is the minimum observed extinction at radii $r < \qty{20}{\kpc}$ with optical recombination lines \citep{2023ApJ...947...33R}. Colors represent \ot/\osix, as in Figure~\ref{fig:o2o6}(b). The model grids show a range of magnetic parameter $b\equiv Bn^{-1/2}$ in units of $\mu$G\,cm$^{-3/2}$ and shock velocity in \kms. LMC models provide the best fit to the observed ratios, though this conclusion could change in the presence of attenuation.}
\label{fig:shock-models}
\end{figure*}
In comparing to the shock models, we also consider a correction for attenuation. The optical recombination lines are consistent with a range of attenuation values $\ebv=0.3-1.0$ mag at $r \la \qty{20}{\kpc}$, and no attenuation at larger radius \citep{2023ApJ...947...33R}. If these apply also to the observed UV lines, then even small amounts of attenuation can change the ratios in Figure~\ref{fig:shock-models}. Using the \texttt{dust\_extinction} package \citep{Gordon2024}, we display the magnitude of a fiducial correction using the \citet{2023ApJ...950...86G} models, $R_\mathrm{V} = 3.1$, and a modest $\ebv = 0.3$ mag of attenuation. This would apply only to bins with $r < \qty{20}{\kpc}$. We caution that the extinction correction is quite uncertain, and this is exercise is intended mainly to illustrate the possible impact of extinction.

The Solar abundance, $n = \qty{1}{\per\cm\cubed}$ model does not overlap the data, as it does for the optical ratios in the outer wind \citep{2023ApJ...947...33R}. A lower, LMC-like abundance is potentially consistent with the data if there is no extinction. A lower abundance may be consistent with absorbers in the CGM, which point to $\langle Z \rangle = Z_\odot/3$ at $z \approx 0.2$ \citep{2017ApJ...837..169P}. The inner, $r < \qty{30}{\kpc}$ bins overlap with the 400~\kms\ models, which are the best fit to the optical line ratios in this part of the nebula \citep{2023ApJ...947...33R}. The outer bins are consistent with 300~\kms, while the best-fit models in the optical are closer to $\approx$200~\kms. However, in the presence of extinction, the inner points move outside of the model grids. Increasing the density to \qty{100}{\per\cubic\cm}, rather than decreasing the abundances, yields a match to the inner, attenuation-corrected data. However, the observed data (uncorrected for attenuation) are then only compatible with shock velocity \qty{>500}{\km\per\second}.

In summary, the lower-metallicity shock models are the best overall match to the data, but only in the absence of extinction. It is conceivable that \ot\ and \osix\ experience different levels of extinction, particularly if they arise at different layers within a cloud in the shock. For instance, if \ot\ arises further into the cloud interior, then it could experience higher extinction. Correcting for this will raise, rather than lower, the \ot/\osix\ ratio. The attenuation correction is also sensitive to the exact values of \ebv, which has limited spatial constraints; the choice of extinction curve; and the choice of $R_V$. There may also be differences arising due to large-scale geometry (see below).

A second uncertainty is the susceptibility of \lya\ to radiative transfer effects. The shock models assume the \lya\ line is emitted from the same volume as other lines. However, \lya\ is {emitted from starburst production sites in the galaxy and, if it escapes, scatters off the wind}. In J1156, the \lya\ nebula is much more compact than the \osix\ halo, and may arise due to scattering \citep{2016ApJ...828...49H}. In contrast, we show above that the data for Makani are consistent with a \lya\ nebula of the same morphology as \osix. If \lya\ is at least in part due to scattering, then removing this contribution from the data would raise the \osix/\lya\ ratio of emission attributable to in-situ processes, like shock ionization. This would improve the match to the shock models.

% {\bf \lya\ may in fact experience significant escape from the inner arcsecond of production sites in the Makani starburst. At least two data points are consistent with a significant \lya\ escape fraction: lack of \ion{Mg}{2} absorption and optically-thin \ion{Mg}{2} emission; and the low inferred SFR from H$\alpha$ \citep{2019Natur.574..643R, 2023ApJ...947...33R}. The parent sample also shows \ntll/\sutl\ ratios that may be consistent with high $f_\mathrm{esc}^\lya$ \citep{2021ApJ...923..275P}.}

For comparison, \osix\ emission has been observed spectroscopically in the AGN-driven outflow surrounding a $z=0.123$ quasar \citep{2020ApJ...890L..28S}. The ionization of the gas in the nucleus and at 10~kpc radius produces $\osix/\mathrm{Ly}\alpha = 0.2$, which is identical to what we observe in the Makani wind. In this case, the wind is AGN-photoionized, which is not the case for Makani, since it has no detectable AGN \citep{2014MNRAS.441.3417S, 2019Natur.574..643R}.

If the UV lines require an ionization mechanism other than shocks, then the morphology of the line ratios offer a clue to their origin. The map of \ot/\osix\ (Figure~\ref{fig:o2o6}(a)) shows that the highest values are found in the East and West high surface-brightness regions of the nebula, at the bases of the lobes of the hourglass shape. The lowest values are found directly North and South of the nucleus and in the most extended regions of the nebula. In a scenario where the hourglass is a container for a hot wind fluid that propagates primarily North and South, then the lower values of \ot/\osix\ along this ``channel'' could trace ionization of the gas through interaction with this fluid \citep{2001ApJ...554.1021H, 2018MNRAS.474.1688C}. As an example of another ionization mechanism, models of turbulent mixing layers in cloud/wind interfaces predict $\ot/\osix\ \ga 10$ \citep{1993ApJ...407...83S, 2023ApJ...950...91C}. These values are significantly higher than observed, but could be compatible with the data if the \ot\ lines are preferentially attenuated. Finally, flickering AGN can also ionize \osix, which then recombines within 10$^7$\;yr, well after lower-ionization species \citep{2018MNRAS.474.4740O}.

Alternatively, these morphological differences may represent differences in extinction. If the East and West regions have higher extinction---as possibly indicated by the dense gas observed in these directions through CO measurements \citep{2019Natur.574..643R}---then \osix\ would be suppressed in these regions compared to \ot, raising the observed \ot/\osix. As pointed out by \citet{2024AJ....168...11K}, this difference in optical depth may impact which part of the wind is probed by each line, if it also reflects a difference in geometric depth.

%photoionization vs. collisional ionization \citep{2024ApJ...962...15Z}
%ionization in disk-halo interface \citep{2024arXiv240902182D}
%\osix\ emission in M82 superwind \citep{2024AJ....168...11K}
%the dominant role of feedback in the inner CGM \citep{2020ApJ...903...32F}
%hot phase mass loading factor

\subsection{Mass and Density}

The typical mass of \osix\ in the CGM of a galaxy with $M_* = 10^{11}$~\msun\ is $M_\mathrm{OVI} \approx 3\times10^6$~\msun\ \citep{2022ApJ...927..147T}. Using this ensemble measurement---and making the assumption that the \osix\ observed in Makani dominates the estimated mass budget of \osix\ within the virial radius---we estimate the density in the \osix\ nebula of Makani. The \osix\ volume emission rate is equal to the volumetric cooling rate in \osix, $\Lambda_\mathrm{OVI}$ \citep{1993ApJS...88..253S}. Following \citet{2016ApJ...828...49H}, we adopt the Solar metallicity value of $\Lambda_\mathrm{OVI}$ from \citet{2010MNRAS.408.1120B}, $\Lambda_\mathrm{OVI}^{Z_\odot} = (\qty{5e-23}{\erg\per\second\cm\cubed})\,n_\mathrm{e}^{2}$, and scale by $Z$: $\Lambda_\mathrm{OVI} =\Lambda_\mathrm{OVI}^{Z_\odot}\,(Z/Z_\odot)$. The mass and luminosity are then related as
\begin{align}
M_\mathrm{OVI} & = \rho_\mathrm{OVI}\,L_\mathrm{OVI}\,\Lambda_\mathrm{OVI}^{-1}\\
& = 16 m_\mathrm{p}\,n_\mathrm{e}\,(N_\mathrm{O}/N_\mathrm{H})\,L_\mathrm{OVI}[\Lambda_\mathrm{OVI}^{Z_\odot}\,(Z/Z_\odot)]^{-1}\\
& = (1.32\times10^5\msun) \left( \frac{L_\mathrm{OVI}}{10^{42}\,\mathrm{erg}\,\mathrm{s}^{-1}}\right) \left( \frac{\qty{1}{\per\cm\cubed}}{n_\mathrm{e}}\right) \left( \frac{Z_\odot}{Z}\right) \label{eq:masslum}
\end{align}
Here, $\rho_\mathrm{OVI}$ is the mass density of \osix\ atoms. We assume that the plasma is completely ionized and that \osix\ dominates the oxygen budget during the $10^{5.5}$~K phase, which means $n_\mathrm{OVI} \approx n_\mathrm{e}\,(N_\mathrm{O}/N_\mathrm{H})$. We adopt the Solar oxygen abundance of \citet{2021A&A...653A.141A}.

From Eq.~\ref{eq:masslum}, the observed \osix\ luminosity in Makani is then equivalent to the ensemble halo mass at a density of $n_\mathrm{e} = \qty{0.5}{\per\cm\cubed}$, assuming a metallicity of $Z_\odot/3$ \citep{2017ApJ...837..169P}. This value is identical to that inferred for the J1156 nebula \citep{2016ApJ...828...49H}. It is similar to the density of the warm, ionized, pre-shock gas in the outer parts of the Makani wind, as inferred from shock models \citep{2023ApJ...947...33R}. The pre-shock density inferred for the inner wind is about 10$\times$ higher. Correcting for possible extinction would raise the inferred density proportionally.

Though we do not measure density $n$ or column density $N$ directly, our estimates for the two are roughly consistent. Since luminosity $L\propto n^2V$ and $N \propto nL$---where $V$ is the emitting volume and $L$ is the path length---then $L/N \propto nV/L \sim nA$, where $A$ is the area of a cross-section of the emitting volume. Comparing to J1156, $L$ in Makani is 20$\times$ higher (Section~\ref{sec:wind-prop}) while our $N$ estimate is only 4$\times$ bigger (Section~\ref{sec:scattering}), so that $(L/N)_\mathrm{Makani} = 5(L/N)_\mathrm{J1156}$. The estimated densities of the two sources are the same, while the area $A$ of Makani's nebula is larger than that of J1156 by the ratio of their half-light radii squared, $(20/12.6)^2\sim2.5$. Thus, $(nA)_\mathrm{Makani} = 2.5 (nA)_\mathrm{J1156}$. The two sides of the equation are thus consistent within a factor of 2, which is reasonable given the uncertainties and assumptions.

Finally, we note that the cooling time of \osix\ is $t_\mathrm{cool}\approx1$~Myr \citep{2007ApJS..168..213G}. If the ensemble \osix\ mass applies to Makani, and \osix\ represents gas condensing from the hot to the cold phase, then the predicted total oxygen cooling rate in the Makani wind is $\dot{M}^\mathrm{OVI}_\mathrm{cool} = M_\mathrm{OVI}/t_\mathrm{cool} \approx 3\,\smpy$. The total predicted cooling rate in the nebula is then $\dot{M}_\mathrm{cool} \approx 1500$~\smpy, for $Z = Z_\odot/3$. If a third of this cooling in the inner wind, scaling by the relative flux of the inner exponential, then the cooling rate at $r \la \qty{15}{\kpc}$ is comparable to the mass outflow rate of the fast, inner wind \citep{2023ApJ...947...33R}. However, the remaining cooling rate is two orders of magnitude greater than that in the slow, outer wind \citep{2023ApJ...947...33R}. This points to the possibility of the increased importance of cooling vs. mass outflow for the cloud mass budget as the wind ages, slows, and moves to larger radius.

\section{Conclusions}

The galaxy Makani is a compact, massive ($M_* \sim 10^{11}$~\msun) galaxy observed at a lookback time of 4.6~Gyr \citep{2014MNRAS.441.3417S}.  It hosts a \qty{100}{\kpc} outflow detected in \ot\ using ground-based integral field spectroscopy with KCWI \citep{2019Natur.574..643R}. This $\sim$ 10$^{10}$~\msun\ outflow has been produced over 400~Myr through two strong starburst events \citep{2019Natur.574..643R}. The warm, ionized line-emitting nebula is consistent with shock excitation and powering by a momentum-conserving flow from the starburst \citep{2023ApJ...947...33R}.

Using the narrowband filters of the ACS Solar Blind Channel on \hst, we have performed 20 orbits of deep imaging to produce F150LP and F165LP images of Makani. We have shown that there is significant, extended emission in F150LP and F165LP that arise primarily from \osix$+$\lya\ and \lya\ emission, respectively. This extended emission is significantly detected at radii \qtyrange{\approx1.5}{5}{\arcsecond} in F165LP and \qtyrange{\approx1.5}{8.5}{\arcsecond} in F150LP.  Based on an analysis of the noise in the F165LP image, we infer that it is also consistent with the more extended emission observed in F150LP. This corresponds to a UV line-emitting nebula over projected radii of \qtyrange{10}{50}{\kpc}.

Using Voronoi binning and the F165LP image to separate \osix\ and \lya, we produce an \osix\ image that strongly resembles the hourglass morphology of the \ot\ emission observed in Makani. The \ot/\osix\ and \osix/\lya\ line ratios are potentially consistent with shock models \citep{2008ApJS..178...20A}, with the caveats that attenuation and radiative transfer effects on \lya\ are uncertain. Turbulent mixing layer models \citep[e.g.][]{2023ApJ...950...91C} predict \ot/\osix\ ratios higher than observed. However, the spatial coincidence of the \ot\ and \osix\ gas strongly implies that the latter arises in the interface between a hot, $10^7$~K phase---the ambient CGM and/or the fluid driving the wind---and cool clouds. The \osix\ arises as gas cools from a hot phase, perhaps contributing to in-situ cloud growth. The morphology of the \osix\ outflow is consistent with models and simulations of gas cooling in a hot wind.

The halo or CGM of only one other galaxy has been imaged in \osix. Makani has a star formation rate 10$\times$ that of J1156 \citep{2016ApJ...828...49H}, and its $L_\mathrm{OVI}\approx\qty{4e42}{\erg\per\second}$ is correspondingly larger. The \osix\ emission in Makani extends $\ga2\times$ as far as that J1156, and has a double-exponential---rather than single-exponential in J1156---radial profile. Nonetheless, the estimated density in the \osix-emitting phase, $n_\mathrm{e}\approx\qty{0.5}{\per\cm\cubed}$, may be similar. Spectroscopy of \osix\ in Makani would, in combination with the imaging data, better constrain the density and inner structure of the wind, as well as the kinematics of the coronal gas and possible scattering geometries.

This experiment represents the bleeding edge of the UV imaging capability of \hst\ for detecting emission in the CGM of galaxies. A future, large-area, UV space-based mission with narrowband imaging capability or integral field spectroscopy---such as the Habitable World Observatory---will be poised to detect such signals in larger samples of galaxies \citep{2019arXiv191206219T}. Along with the previous detection in J1156, we have shown that detecting cooling gas, which primarily emits in UV line emission \citep{2013MNRAS.430.3292B}, may be easier than expected if galactic winds enhances the detectability of such lines. Natural future targets might be the ubiquitous \ot\ nebula surrounding $z < 1$ quasars \citep{2024ApJ...966..218J}.

\begin{acknowledgments}
We thank Sally Heap for the idea to apply the differential narrowband technique to Makani. We thank Matthew Hayes for sharing data on J1156 and for other helpful conversations on the analysis. This research is based on observations made with the NASA/ESA Hubble Space Telescope obtained from the Space Telescope Science Institute, which is operated by the Association of Universities for Research in Astronomy, Inc., under NASA contract NAS 5–26555. These observations are associated with program \#16231. Support for program \#16231 was provided by NASA through a grant from the Space Telescope Science Institute.
\end{acknowledgments}

\vspace{5mm}
\facilities{HST(ACS-SBC)}

\software{\texttt{astropy} \citep{2013A&A...558A..33A,2018AJ....156..123A,2022ApJ...935..167A}, \texttt{DrizzlePac} \citep{2012ascl.soft12011S}, \texttt{dust\_extinction} \citep{Gordon2024}, \texttt{photutils} \citep{larry_bradley_2024_13989456}, \texttt{regions} \citep{larry_bradley_2024_13852178}, \texttt{reproject} \citep{2020ascl.soft11023R}, \texttt{stsynphot} \citep{2020ascl.soft10003S}, \texttt{synphot} \citep{2018ascl.soft11001S}, \texttt{VorBin} \citep{2003MNRAS.342..345C}}

\bibliography{makani-ovi}{}
\bibliographystyle{aasjournal}

\end{document}